\documentclass{article}
\usepackage{amssymb,axodraw,cite,graphicx}

\title{Effective Gauge Theories, The Renormalization Group, and
High--T${}_{\rm c}$ Superconductivity. \thanks{Based on lectures given
by N.E.M. at the XXXVIII Cracow School of Theoretical Physics, {\sl
New Quantum Phases, Elementary Excitations and Renormalization in High
Energy and Condensed Matter Physics}; Zakopane, Poland; June 1--10,
1998.}}

\author{A. Campbell--Smith and N.E. Mavromatos. \\ { \small
\sl Theoretical Physics, (University of Oxford),1 Keble Road, Oxford,
OX1 3NP, U.K.}}

\date{ }

\newcommand{\be}{\begin{equation}}
\newcommand{\ee}{\end{equation}}
\newcommand{\ba}{\begin{eqnarray}}
\newcommand{\ea}{\end{eqnarray}}

\newcommand{\dsl}{\not\!\partial}
\newcommand{\asl}{\not \! a}
\newcommand{\psl}{\not \; \!\!\! p}

\begin{document}

\maketitle

\begin{abstract}
These lectures serve as an introduction to the renormalization group
approach to effective field theories, with emphasis on systems with a
Fermi surface.  For such systems, demanding appropriate scaling with
respect to the renormalization group for the appropriate excitations
leads directly to the important concept of quasiparticles and the
connexion between large--$N_{\rm f}$ treatments and renormalization
group running in theory space.  In such treatments $N_{\rm f}$ denotes
the number of effective fermionic degrees of freedom above the Fermi
surface; this number is roughly proportional to the size of
the Fermi surface.  As an application of these ideas, non--trivial
infra red structure in three dimensional U(1) gauge theory is
discussed, along with applications to the normal phase physics of
high--T$_{\rm c}$ superconductors, in an attempt to explain the
experimentally observed deviations from Fermi liquid behaviour.
Specifically, the direct current resistivity of the theory is computed
at finite temperatures, $T$, and is found to acquire ${\mathcal
O}(1/N_{\rm f})$ corrections to the linear $T$ behaviour.  Such
scaling corrections are consistent with recent experimental
observations in high T${}_{\rm c}$ superconducting cuprates.
\end{abstract}

\vspace*{-11cm}
\begin{flushright}
OUTP--98--73P\\
cond--mat/9810324
\end{flushright}
\vspace*{14cm}

\newpage

\tableofcontents
\addtocontents{nick.toc}

\newpage

\section{Introduction.}
\setcounter{equation}{0}

The concept of effective field theory
\cite{wilson75,polch:rg:tasi,shankar90,benfatto90} is an old one, and
has wide applications in physics, ranging from condensed matter to
high energy physics and cosmology.  The effective field theory method
isolates properly those degrees of freedom in a dynamical system which
are driving the dynamics in a certain range of energy--momentum.

Formally, an effective field theory is represented by a functional
integral in which we perform an appropriate splitting of the
field--theoretic degrees of freedom $\{ \phi(x) \}$:
\begin{equation} \label{split}
\{ \phi(x) \} = \{ \phi_{\rm H} (x) \} + \{ \phi_{\rm L} (x) \}.
\end{equation}
The fields $\phi_{\rm H}$ ($\phi_{\rm L}$) have Fourier components
corresponding to momenta $k>\Lambda$ ($k<\Lambda$) where $\Lambda$ is
a characteristic energy scale in the problem which {\em defines} the
low energy effective dynamics.  Now the effective field theory below
the scale $\Lambda$ is obtained formally by integrating out the
high--frequency modes $\phi_{\rm H}$ in the functional integral.  What
remains after the integrations defines the {\em effective Lagrangian
density} $\mathcal L_{\rm eff} (\phi_{\rm L} , \dot{\phi}_{\rm L} )$
which is an infinite series in a derivative expansion and describes
the low energy dynamics.  The issue of whether the splitting
(\ref{split}) can always lead to a meaningful effective theory is a
very complicated one, and in general depends on the details of the
dynamics and the cut--off scale $\Lambda$.  In most cases of interest,
however, the concept of an effective theory is useful in describing
the basic features of the underlying dynamics (below the scale
$\Lambda$) in a simple way.

One of the most important tools in the study of effective field
theories is the renormalization group
\cite{wilson75,kadanoff71,wilson72,wilson+kogut}.  This technique
allows one to group effective field theories with apparently very
different interactions into categories which emphasise their common
features, {e.g.}\ similar scaling exponents of certain correlation
functions (which can be measured experimentally).  Such groups of
effective theories are termed ``universality classes'' and prove
essential in understanding the similar properties of ostensibly
different physical systems: all systems in the same universality class
flow to the same renormalization group fixed point \cite{wilson75}.

The aim of this set of lectures is to introduce the concept of an
effective field theory and the associated renormalization group
techniques, first in a generic way, and later on through some specific
and physically interesting examples; these will include the BCS
superconductivity phenomenon (as a guided exercise for the reader) and
a possible explanation of the abnormal properties of high--temperature
superconductors in their normal phase.

The structure of the lectures is as follows:  in the first half of the
lectures (sections \ref{rgsec}--\ref{rg+sf}) the basic features of the
renormalization group approach to effective field theories is
discussed briefly.  Particular attention is paid to discussing
scaling properties of systems with a Fermi surface, relevant for
condensed matter applications.  The important concept of
quasiparticles is introduced in section \ref{rg+sf}.  There are
excitations of the Fermi surface which are appropriately dressed so as
to have the correct scaling under the renormalization group.

An interesting application concerning the representation of the Landau
Fermi liquid theory as a theory with a trivial infra red fixed point
concludes the first half of the lectures.  In this model the
quasiparticle degrees of freedom below a scale $\Lambda$ appear as
fermions with a $\Lambda$--dependent ``flavour number'' $N_{\rm f}(\Lambda)$,
which runs to infinity as $\Lambda/P_{\rm F}\longrightarrow0$ ($P_{\rm
F}$ being the typical size of the Fermi surface).

In the second half of the lectures (section \ref{qed3} ff.) a second
application of the renormalization group and effective theories is
considered:  that of trying to understand the abnormal behaviour of
high temperature superconductors in their normal (chirally symmetric)
phase in terms of deviations from the Fermi liquid trivial infra red
fixed point.  In the context of the gauge theory approach to the
physics of doped antiferromagnets, believed to simulate the physics of
high temperature superconductivity, it is demonstrated using methods of
effective field theory that three dimensional U(1) gauge theory
(QED${}_3$) is characterized by a non--trivial infra red fixed point.
The fermion--gauge field interaction vertex becomes marginally
relevant and drives the theory to non--Fermi liquid behaviour in the
infra red.  The fixed point structure is non--perturbative and is
discovered using a renormalization group improved Dyson--Schwinger
analysis \cite{aitch+mav:prb}.

The interesting feature of the QED${}_3$ Dyson--Schwinger resummed
problem is that the ``running'' coupling coincides with the inverse of
a fermion flavour number, thereby providing an interesting application
of the above--mentioned effective running of flavour number in the
context of condensed matter physics.  Some physical consequences of
this phenomenon, concerning the scaling of the electrical resistivity
in QED${}_3$ at finite temperature are discussed with the aim of
comparing the results with the phenomenology of the normal phase of
high temperature superconductors (section \ref{last}).

Finally, we present conclusions and outlook in section \ref{conc}.

\section{Lecture I (i):  Effective Field Theory.} \label{rgsec}
\setcounter{equation}{0}

\subsection{Generic analysis of Wilsonian approach.}

Consider a field theory in which an energy scale $E_0$ is introduced.
This energy scale need not be associated with the normal
divergence--cancelling cut--offs in field theory.  Effective field
theory is a method for analysing the physics at lower energy scales
$E\ll E_0$.  To construct the effective field theory, split the fields
into high and low frequency components $\phi_{\rm H}$ and $\phi_{\rm
L}$ with frequencies above and below the scale $\Lambda \sim E_0 /
\hbar$ respectively (from here on $\hbar\equiv 1$):
\begin{equation}
\{ \phi(\omega) \} = \{ \phi_{\rm H} (\omega) \} + \{ \phi_{\rm L}
(\omega) \}
\end{equation}
where
\begin{eqnarray}
\{ \phi_{\rm H} (\omega) \} &=& \{ \phi(\omega) \; : \; \omega>\Lambda \}, \nonumber\\
\{ \phi_{\rm L} (\omega) \} &=& \{ \phi(\omega) \; : \; \omega<\Lambda \};
\end{eqnarray}
in general the split can be performed smoothly or sharply:  the
specifics of the split will not matter here.

Now the high-frequency components $\phi_{\rm H}$ are integrated out in
the functional integral
\begin{equation}
\int D\phi_{\rm L} \int D\phi_{\rm H} \; e^{i S(\phi_{\rm
H},\phi_{\rm L})} = \int D\phi_{\rm L}\; e^{i
S_{\Lambda}(\phi_{\rm L})},
\end{equation}
and the {\em Wilsonian effective action} $S_\Lambda (\phi_{\rm L})$ is
given by
\begin{equation} \label{effac}
e^{i S_\Lambda (\phi_{\rm L})} = \int D\phi_{\rm H}\;
e^{iS(\phi_{\rm L},\phi_{\rm H})}.
\end{equation}

The effective action $S_\Lambda$ can be expanded in a complete set of
local operators:
\begin{equation} \label{opexpansion}
S_\Lambda = S_0 (\Lambda,g^*) + \sum_i \int d^D x \; g^i \Theta_i ,
\end{equation}
where the sum runs over all local operators $\Theta_i$ allowed by the
symmetries of the model and the $\{g^i\}$ are the associated
couplings, which may be thought of as coordinates in coupling space.
The set of couplings $\{g^*\}$ denote the (trivial) fixed point of the
theory (see section \ref{rgflows}), which for convenience can be
chosen to be $g^*=0$ (the origin of coupling space) so that the
expansion point $S_0$ is the free action.  Some elementary dimensional
analysis from the free action yields the {\em scaling dimension} of
the operators and couplings:
\begin{equation}
\begin{array}{cll}
\Theta_i &\longrightarrow& E^{h_i}, \\ g^i
&\longrightarrow& E^{D-h_i}.
\end{array}
\end{equation}
These scalings can be used to derive a dimensional estimate of the
magnitude of an operator in $S_\Lambda$,
\[
\int d^D x\; \Theta_i \sim
E^{h_i -D}.
\]
Introducing the dimensionless couplings
\begin{equation}
\lambda^i = g^i \,\Lambda^{h_i -D},
\end{equation}
it can be seen that the $i^{\rm th}$ term in the action is of order
\begin{equation}
\lambda^i \left( \frac{E}{\Lambda} \right)^{h_i -D}.
\end{equation}

Using this dimensional estimate, the operators in the expansion
(\ref{opexpansion}) can be classified according as the value of $h_i -
D$ is positive, negative or vanishing; see table \ref{operators}.  If
$h_i - D < 0$ the operator becomes more important at lower energies,
and is called {\em relevant}.  An operator with vanishing $h_i -D$ is
equally important at all energies and is called {\em marginal}.
Operators with $h_i -D > 0$ are {\em irrelevant}, for they become less
and less important at low energies.

\begin{table}[tbp] \label{operators}
\begin{center}
\begin{tabular}{|c|c|c|c|}\hline
&&&\\
$h_i -D$ & Size as $E\longrightarrow 0$ & Type of Operator &  Type of
Theory \\ 
&&&\\ \hline
&&&\\
$< 0$ & Grows & Relevant & Super--renormalizable \\
&&&\\
\hline
&&&\\
$0$ & Constant (scale invariant) & Marginal & Strictly renormalizable \\
&&&\\
\hline
&&&\\
$> 0$ & Decays & Irrelevant & Non--renormalizable \\
&&&\\
\hline
\end{tabular}
\caption{Classification of operators in an effective field theory.}
\end{center}
\end{table}

The lesson to be learned from the power counting above is that the low
energy physics is only sensitive to the high energy theory through the
marginal and relevant couplings.  In most cases there is a finite
number of relevant and marginal couplings, so in principle the low
energy physics depends on only a finite number of parameters.  There
are subtleties, however, due to possible infra red structure,
{i.e.}\ divergences in the low energy theory.  The simple power counting
above is done from the free action, and so the interactions present in
the full effective action (\ref{effac}) can affect the behaviour:
operators can change between being marginal, relevant and irrelevant
as a result of the interactions.

\subsection{Renormalization group flow equations:
$\beta$--functions.} \label{rgflows}

The scaling derived from simple power counting is modified by
interactions in the effective theory; these effects are encoded in the
$\beta$--function.  The $\beta$--functions for the renormalized
couplings $g^i (E)$ are defined as follows:
\begin{equation}
\beta^i \equiv E \frac{\partial}{\partial E} g^i (E) = y^i g^i (E) +
C^i_{jk} \, g^j \, g^k + \ldots,
\end{equation}
where $y^i$ are the anomalous dimensions (due to quantum corrections)
and the $C^i_{jk}$ are the coefficients of the operator product
expansion pertaining to the three--point functions of the theory:
\begin{equation}
\langle \Theta_i (x_1) \, \Theta_j (x_2) \, \Theta_k (x_3) \rangle_0 =
C_{ijk} |x_{12}|^{\delta_{ij}-D} \, |x_{13}|^{\delta_{ik}-D} \,
|x_{23}|^{\delta_{jk}-D},
\end{equation}
where $\delta_{ij} = y_i + y_j - y_k$, etc., $|x_{ij}| = |x_i -
x_j|$ and where $\langle \cdots \rangle_0$ indicates correlators taken
with respect to $S_0(\Lambda,g^*)$ (see equation (\ref{opexpansion})).

In coupling space indices are raised and lowered by the so--called
Zamolodchikov \cite{zamolodchikov86} metric:
\begin{equation}
G_{ij} = |x|^{2D-y_i -y_j} \, \langle \Theta_i (x) \, \Theta_j (0)
\rangle_{S_\Lambda}
\end{equation}
where now the symbol $\langle \cdots \rangle_{S_\Lambda}$ indicates
correlators taken with respect to the full (interacting) action,
equation (\ref{effac}).  The covariant coefficients $C_{ijk}$
appearing in the three--point functions are totally symmetric in their
indices.

Close to the fixed point (at least to order $g^2$), the
$\beta$--functions defined above are related \cite{mavromatos89} to a
gradient flow in coupling space
\begin{equation} \label{phiflow}
\partial_i \Phi (g,g^*,\Lambda) = G_{ij} (g^*,\Lambda) \beta^j (g, g^*,\Lambda).
\end{equation}
The renormalization group invariant flow function $\Phi$ in two
dimensional systems has been related to the components of the stress
tensor of the theory \cite{zamolodchikov86}.  In higher dimensions
$\Phi$ is still not known in closed form, although attempts have been
made to relate it to stress tensor components by appropriate extension
of the two dimensional case \cite{forte98}.  Notice that the total
symmetry of $C_{ijk}$ is crucial \cite{mavromatos89} for the gradient
flow (\ref{phiflow}).

In the case of two dimensional unitary theories the metric $G_{ij}$ is
manifestly positive definite \cite{zamolodchikov86} and hence relations
like equation (\ref{phiflow}) imply, in view of the renormalization
group invariance of $\Phi$, that the flow function decreases along the
renormalization group trajectories in coupling space:
\begin{equation} \label{flowlesszero}
\partial_t \Phi = - \beta^i \partial_i \Phi = - \beta^i G_{ij} \beta^j
< 0,
\end{equation}
where $t = \ln (E/\Lambda)$ is a renormalization group scale.
According to reference \cite{zamolodchikov86} the value of the flow
function at a fixed point coincides with the central charge of the
corresponding two dimensional conformal field theory.  Hence the
relation (\ref{flowlesszero}) implies that under relevant
perturbations a unitary theory flows along a direction of decreasing
central charge.  This poses interesting restrictions on the flow of
two dimensional theories which might have interesting physical
applications \cite{ludwig87}.  In higher dimensions the proof of such
an irreversibility is not yet complete but recently there have been
interesting attempts \cite{forte98}.  Such a theorem on the
irreversibility of the renormalization group flow in $D$ dimensions is
expected to hold on general grounds for effective field theories are
plagued by loss of information from the modes above the scale
$\Lambda$ which are integrated out.  Such modes contribute a non--zero
entropy change which in the case of two dimensional physics has been
shown to correspond to the flow function $\Phi$.

\subsubsection{Example:  a single marginal coupling.}

A single marginal coupling will typically have a $\beta$--function as
follows:
\begin{equation}
\beta^g = b g^2 + {\mathcal O}(g^3) ,
\end{equation}
which can readily be integrated to give the solution
\begin{equation}
g(E) = \frac{g(\Lambda)}{1+ bg(\Lambda) \ln\left( \frac{\Lambda}{E}
\right)}.
\end{equation}
For $b>0$ the coupling decreases at low energies and is marginally
irrelevant.  If $b<0$ the coupling grows and is marginally relevant.
A strictly marginal coupling is only obtained if the $\beta$--function
vanishes to all orders in $g$.  So it is seen that the physics of a
system with a marginal coupling depends on the details of the problem,
here encoded in $b$.  A marginally relevant coupling can lead to
interesting effects; for example, in QCD a marginally relevant
coupling leads to confinement and chiral symmetry breaking; similarly
in condensed matter models of the resistivity in the presence of
magnetic impurities, the Kondo effect \cite{kondo64} leads to an
increase in the resistivity in the deep infra red (the very low
temperature region).

For a single coupling the gradient flow relation (\ref{phiflow}) is
trivial to prove.  It is not at all trivial to prove the relation for
models with more than one coupling, for the existence of a gradient
flow requires the curl--free condition on the quantity $G_{ij}
\beta^j$.  This has been shown in $D$ dimensions at least in the
neighbourhood of a fixed point (to order $g^2$) in reference
\cite{mavromatos89}.

\subsubsection{Super--renormalizable (relevant) couplings: naturalness.}

In effective field theories, non--renormalizable terms do not cause
problems, for there is an ultra violet cut--off which automatically
acts as a regulator for the associated divergences.  Indeed,
non--renormalizable terms {\em have} to appear at some scale in the
effective theory: in fact the information about where the cut--off
must lie is encoded in these terms.  For effective field theories
there is a new problem, not found in conventional field theories:
super--renormalizable terms, which grow below the cut--off.  This
means that, without unnatural fine--tuning of the original parameters,
all masses in the effective field theory must be of order of the
cut--off; this is a contradiction, however, for such fields cannot
appear in the effective theory at all.  So effective theories must be
{\em natural}, meaning that there must exist symmetries which force
the masses to vanish.  A pertinent example of this is the U${}_{\rm
S}$(1) effective gauge field theory of doped antiferromagnets at the
$d$-wave gap (see section \ref{su2crossu1}): the charged excitations
are described by Dirac fermions whose masses are forced to vanish by
chiral symmetry; there are also gauge fields forced massless by gauge
invariance.  To include scalars in a model their masses must be
forbidden by Goldstone's theorem or supersymmetry.

There is also a problem with super--renormalizable interactions: at
scales below the cut--off, the (dimensionless) coupling will grow to
larger than unity, and hence the low energy theory may be described by
new degrees of freedom: bound states, condensates, etc.  An exception
to this is when the infra red behaviour is governed by a
(non--trivial) fixed point, in which case the theory may still be
described by the original degrees of freedom.  An example is QED${}_3$
(see section \ref{qed3}), which is super--renormalizable, but
renormalization group improved Dyson--Schwinger analysis reveals a
non--trivial infra red fixed point \cite{aitch+mav:prb}.  As will be
discussed later in lecture II (section \ref{qed3} ff.) this is
relevant for the normal phase of high--T${}_{\rm c}$ superconductors.

\subsection{Concluding remarks.}

\begin{itemize}
\item Effective field theory is a tool for understanding the physics
of renormalization.
\item Interactions in an effective field theory can be classified as
{\em relevant, marginal, or irrelevant} according to their scaling
properties.
\item The infra red physics is only sensitive to the high energy
theory through the marginal and relevant couplings.
\item When a marginal coupling grows large, interesting physics can
result.
\item Interesting physics can also arise when a super--renormalizable
coupling is driven to a non--trivial infra red fixed point.
\item Naturalness: effective theories must be {\em natural}, in that
there must exist symmetries which force parameters which would
naturally appear at ${\mathcal O} (\Lambda)$ to vanish.
\end{itemize}

\section{Lecture I (ii): The Renormalization Group, Fermions And The
Fermi Surface.} \label{rg+sf}
\setcounter{equation}{0}

\subsection{Generic analysis.}

Now the renormalization group approach is applied to the theory of the
Fermi surface in condensed matter systems.  This will demonstrate the
need for the concept of quasiparticles and the provide motivation for
large--$N_{\rm f}$ treatments ($N_{\rm f} = $ flavour number)
\cite{benfatto90}.

Slice momentum space into the following sets:  $I_1,\ I_0,\ I_{-1},\
I_{-2}, \ \ldots$

\begin{eqnarray}
I_1 &=& \left\{ \left( k_0,\vec{k} \right)\: : \: k_0^2 + \left[ \frac{
(\vec{k}^2 - P_{\rm F}^2)}{2m} \right]^2 \geqslant P_0^2 \right\} \nonumber\\
I_n &=& \left\{ \left( k_0,\vec{k} \right) \: : \: (2^{n-1} P_0)^2
\leqslant k_0^2 + \left[ \frac{ ( \vec{k}^2 - P_{\rm F}^2 ) }{2m}
\right]^2 \leqslant (2^n P_0)^2 \right\} \qquad \qquad n \leqslant 0.
\end{eqnarray}
The scale $P_0$ is arbitrary, {e.g.}\ the inverse of the range of the
potential.  The Fermi sphere is defined by:  $(k_0, |\vec{k}|=P_{\rm
F})$, and $(2^n P_0)$ is a measure of the distance of the n${}^{\rm th}$ layer
from the Fermi surface.

The fermion propagator is given by \cite{luttinger60}
\begin{equation}
G(t,\vec{x}) = \sum_{n=-\infty}^{n=1} G^{(n)} (t, \vec{x}) \equiv
\sum_{n=-\infty}^{n=1} \int_{I_n} \frac{dk_0 \, d^d k}{(2\pi)^{d+1}}
\frac{e^{-i (k_0 t + \vec{k}\cdot\vec{x} )}}{-ik_0 + (\vec{k}^2 -
P_{\rm F}^2)/2m} .
\end{equation}
The $n^{\rm th}$ summand $G^{(n)}$ is the contribution to the
propagator coming from the layers at a distance ${\mathcal O}(2^n
P_0)$ from the Fermi surface.  This decomposition of the propagator
generates a representation for the fermionic fields:
\begin{equation}
\Psi^{\pm}_{\vec{x}} = \sum_{n=-\infty}^{n=1} \Psi^{(n)\pm}_{\vec{x}},
\end{equation}
where the plus and minus signs are respectively for ``particles'' and
``holes.''

There is a difficulty with the application of renormalization group
techniques to this non--relativistic field theory.  This is illustrated
by an example in three dimensions \cite{benfatto90}:  for large $n$
and large $|\vec{x}|+|t|$, it can be shown that
\begin{equation}
G^{(n)} (t, \vec{x}) \sim 2^{2n} m\, P_{\rm F} \, P_0 \; \left[ t \frac{\sin \left(
P_{\rm F} |\vec{x}| \right)}{P_{\rm F} |\vec{x}|} + \frac{ m \cos
\left(P_{\rm F} |\vec{x}|\right)}{P_{\rm F}}\right] {\mathcal G}(2^n t
P_0 , 2^n \vec{x} P_0).
\end{equation}
While the function ${\mathcal G}$ scales normally, {i.e.}\ depends on
$(t,\vec{x})$ only as $(2^n P_0\, t, 2^n P_0\, \vec{x})$, there is a
problematic oscillation on a scale $P_{\rm F}^{-1}$ and a scaleless
singularity $|\vec{x}|^{-\!1}$ so overall the propagator does not scale
properly.  This is unlike the situation in relativistic field theories
as discussed in the section \ref{rgsec} where $P_{\rm F}$ is
vanishing; so how can fields be assigned scaling dimensions and how
can couplings be identified as marginal, relevant or irrelevant?  The
resolution is in the concept of quasiparticles \cite{benfatto90}, which
are the subject of the next subsection.

\subsection{Quasiparticles.} \label{quasi}

Consider the following expansion of the particle/hole fields:
\begin{equation}
\Psi_{\vec{x}}^\pm = \int_{|\vec{\omega}|=1} d\vec{\omega}\; e^{\pm i
P_{\rm F} \vec{\omega}\cdot\vec{x}} \Psi_{\vec{x},\vec{\omega}}^\pm .
\end{equation}
The integration is over the unit sphere, and so $P_{\rm F} \,
\vec{\omega}$ is a momentum on the Fermi sphere.

\begin{figure}
\begin{center}
\begin{picture}(100,100)(0,0)
\PhotonArc(50,50)(40,290,17){-2.6}{2.5}
\PhotonArc(50,50)(40,17,79){1}{1.5}
\PhotonArc(50,50)(40,79,158){-1.4}{2.5}
\PhotonArc(50,50)(40,158,290){2.5}{1.5}
\LongArrow(50,50)(83,69)
\LongArrow(50,50)(82,90)
\DashArrowLine(86,70)(84,91){1}
\Text(70,50)[]{$P_{\rm F}\vec{\omega}$}
\Text(60,70)[r]{$\vec{k}$}
\Text(90,80)[l]{$\vec{\kappa}$}
\end{picture}
\caption{The Fermi Surface of effective radius ${\mathcal O}(P_{\rm
F})$ (assuming a near--spherical shape), showing the quasiparticle
momentum as measured from the surface.\label{fermisurf}}
\end{center}
\end{figure}
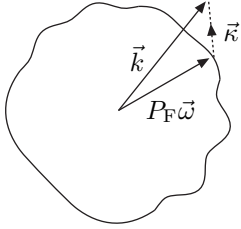

Inserting the Fourier transform of $\Psi_{\vec{x},\vec{\omega}}^\pm$,
\begin{equation}
\Psi_{\vec{x}}^\pm = \int d\vec{k} \int_{|\vec{\omega}|=1}
d\vec{\omega} \; e^{ \pm i ( P_{\rm F} \vec{\omega} -
\vec{k})\cdot\vec{x}} \tilde{\Psi}_{\vec{x},\vec{\omega}}^\pm 
\end{equation}
So here the quantity $-\vec{\kappa} = P_{\rm F}\,\vec{\omega} -
\vec{k}$ plays the r\^{o}le of a momentum, as measured from the
surface of the Fermi surface (see figure \ref{fermisurf}).  The
integration over the orientation $\vec{\omega}$ has to be performed at
the very end of the computations.

The advantage of using the quasiparticle degrees of freedom
was indicated in the last section; it is because they have the correct
scaling behaviour and so admit a conventional renormalization group
treatment.  The quasiparticle propagator reads:
\begin{equation}
G^{(n)} (\vec{x},\vec{\omega}; \vec{x}^{\prime},\vec{\omega}^{\prime})
= \delta(\vec{\omega} -\vec{\omega}^\prime) \, G^{(n)}
(\vec{x}-\vec{x}^\prime ; \vec{\omega} ).
\end{equation}
Now for large $n$ it is possible to show \cite{benfatto90} that the
quasiparticles scale in all dimensions $(D)$ like $2^{n/2}$ and hence have
mass dimension $\frac{1}{2}$:
\begin{equation}
G^{(n)} (\vec{x},\vec{\omega}) \sim 2^n P_{\rm F}^{D-1} \, P_0 \left(
2^n P_0 t- 2iP_0 \vec{\omega}\cdot \vec{x} \right) G^{(n)} ( 2^n P_0
\vec{x} ).
\end{equation}

\begin{figure}
\begin{center}
\begin{picture}(100,100)(0,0)
\CArc(50,50)(40,0,360)
\Line(50,50)(82.17,76.99)
\Line(50,50)(71,86.37)
\CArc(50,50)(18,40,60)
\CArc(50,50)(40.5,40,60)
\CArc(50,50)(41,40,60)
\CArc(50,50)(39,40,60)
\CArc(50,50)(39.5,40,60)
\DashLine(50,50)(92,50){2}
\DashLine(50,50)(89.47,64.36){2}
\DashLine(50,50)(57.29,91.36){2}
\DashLine(50,50)(42.71,91.36){2}
\Text(78,86)[l]{$\triangle\theta\sim {2\Lambda}/{P_{\rm F}}$}
\end{picture}
\caption{The orientation space is sliced into angular cells.  Note
that the width of the cells could in principle be some function of
$\Lambda/P_{\rm F}$.\label{angulars}}
\end{center}
\end{figure}
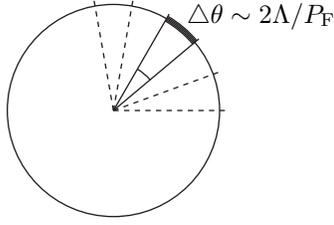

Quasiparticles appear as a consequence of non--trivial Fermi surfaces,
about points at which the dispersion relation has been linearized.  To
illustrate, a model in $2+1$ dimensions is considered, which is
relevant for there is the possibility of applications to the physics
of high--T${}_{\rm c}$ superconductors.  First, slice the orientation
space into angular cells (see figure \ref{angulars}).  For spherical
Fermi surfaces, the angular integration can be replaced by a sum:
\begin{equation}
\int_{|\vec{\omega}|=1} d\vec{\omega} \longrightarrow P_{\rm F}
\int_0^{2\pi} \frac{d\theta}{2\pi} \longrightarrow \sum_i ,
\end{equation}
where the sum is over the angular cells, each of which has width
\begin{equation} \label{cellsize}
\triangle \theta = \frac{2\Lambda}{P_{\rm F}}
\end{equation}
and where $\Lambda$ is some ultra violet cut--off in the theory.  There are
\begin{equation}
N= \frac{2\pi P_{\rm F}}{2\Lambda}
\end{equation}
cells labelled by $i$ \cite{shankar90}.  At each cell the momentum can
be written as follows:
\begin{equation}
\vec{k}_i = P_{\rm F} \vec{\omega}_i + \kappa^{\parallel}_i
\vec{\omega}_i + \kappa^{\perp}_i \vec{t}_i \equiv P_{\rm F}
\vec{\omega}_i + \vec{\kappa}_i .
\end{equation}
The $\kappa^{\parallel}_i$ are the radial components of the momentum,
$\kappa^{\perp}_i$ are the angular displacements from the centre of
the cell and the $\vec{t}_i$ are tangent unit vectors on the Fermi
surface which form a basis for the angular displacements
$\kappa^\perp_i$.  Therefore:
\begin{equation}
\int \frac{d^2 k}{(2\pi)^2} \equiv \int_{-\Lambda}^\Lambda
\frac{dk}{2\pi} P_{\rm F} \int_{-\Lambda/P_{\rm F}}^{\Lambda/P_{\rm
F}} \frac{d\theta}{2\pi} = \int_{-\Lambda}^{\Lambda}
\frac{d\kappa^\parallel}{2\pi} \int_{-\Lambda}^{\Lambda}
\frac{d\kappa^{\perp}}{2\pi} .
\end{equation}
Note that the size of each cell, equation (\ref{cellsize}), need not
be a simple function of $\Lambda/P_{\rm F}$.  More generally each cell
may be allowed to have a size $f(\Lambda/P_{\rm F})$ which can be
determined from the renormalization group scaling; in what follows,
and in section \ref{qed3} for the case of three dimensional U(1) gauge
theory, a connexion will be made between the function $f$ and the
flavour number.

In this $2+1$ dimensional model the kinetic term for free fermions in
the current formalism is as follows:
\begin{equation}
S_0 = \sum_{i=1}^{N} \int \frac{d^2 \kappa_i \,
d\kappa_i^0}{(2\pi)^3}\; \bar{\Psi}_i (\vec{\kappa}_i , \kappa_i^0 )
\left[ i \kappa_i^0 - v^{*} \kappa_i \right] \Psi_i (\vec{\kappa}_i ,
\kappa_i^0 ) .
\end{equation}
The model has the following interaction terms:
\begin{equation}
S_{\rm F} = - \frac{1}{P_{\rm F}} \sum_{i,j = 1}^N \int d{\mathcal F}
\; \bar{\Psi}_j (\vec{\kappa}_4 , \kappa_4^0 ) \, \Psi_j (
\vec{\kappa}_2 , \kappa_2^0 ) \, F_{ij} \, \bar{\Psi}_i
(\vec{\kappa}_3 , \kappa_3^0 ) \, \Psi_i ( \vec{\kappa}_1 , \kappa_1^0
),
\end{equation}
where the measure $d{\mathcal F}$ is
\[
d{\mathcal F} = \left( \prod_{\ell = 1}^4 d^3 \kappa_{\ell}\right)
\delta( \kappa_1^0 + \kappa_2^0 - \kappa_3^0 -\kappa_4^0 )
\,\delta^{(2)} ( \vec{\kappa}_1 + \vec{\kappa}_2 - \vec{\kappa}_3 -
\vec{\kappa}_4 ),
\]
and the coupling is
\[
F_{ij} = F \, \vec{\omega}_i \cdot \vec{\omega}_j .
\]

Now, expressing all the momenta $\vec{\kappa}$ and frequencies $\kappa_0$ in
terms of the cut--off $\Lambda$ it can be seen that the only place
where $\Lambda$ appears is in front of the interaction term, in the
combination $\Lambda/P_{\rm F}$.  This is to be compared with
large--$N_{\rm f}$ models under the replacement:
\begin{equation} \label{Sfarea:N}
f\left( \frac{\Lambda}{P_{\rm F}} \right) \longleftrightarrow
\frac{1}{N_{\rm f}} .
\end{equation}
Here $N_{\rm f}$ can be interpreted as the effective area of the Fermi
surface \cite{shankar94}.  For kinematical reasons the most important
interactions are between $\bar{\Psi} \Psi$ {\em in the same cell}
\cite{polch:rg:tasi}.  In order to allow for maximum momentum transfer
within the framework of the effective theory ({i.e.}\ remaining close
to the Fermi surface) interactions between excitations well separated
on the generic Fermi surface (not exhibiting nesting) are suppressed.
Nesting is of course the case of BCS instabilities, which is the
subject of the exercise in section \ref{exone}.  The infra red limit
of the $2+1$ dimensional model considered ($\Lambda \ll P_{\rm F}$)
corresponds to the $N_{\rm f}\longrightarrow\infty$ limit in the
large--$N_{\rm f}$ model.  Thus the qualitative effect of the
``quasiparticles'' is to increase the effective number of flavours,
where here the flavours are the internal degrees of freedom due to the
Fermi surface.  Note, however, that $N_{\rm f}$ depends on the
cut--off and therefore there is a renormalization group running in
``theory space'' as discussed in sections \ref{qed3} ff.

\subsection{Exercises.} \label{exone}

\subsubsection{BCS Pairing interactions as deviations from Fermi liquid theory.}

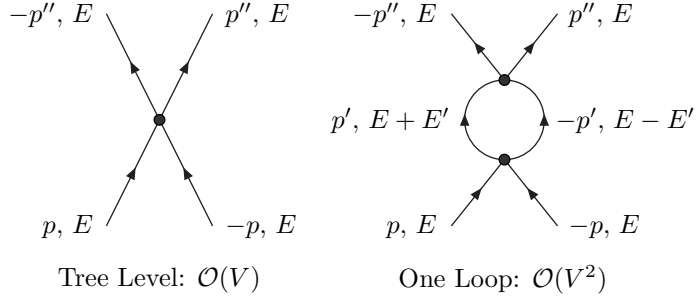
\begin{figure}
\begin{center}
\begin{picture}(220,120)(0,0)
\ArrowLine(10,20)(30,60)
\ArrowLine(30,60)(10,100)
\ArrowLine(30,60)(50,100)
\ArrowLine(50,20)(30,60)
\GCirc(30,60){2}{0.2}
\Text(30,0)[]{Tree Level: ${\mathcal O}(V)$}
\Text(5,20)[r]{$p ,\, E$}
\Text(5,100)[r]{$-p^{\prime\prime} ,\, E$}
\Text(55,20)[l]{$-p ,\, E$}
\Text(55,100)[l]{$p^{\prime\prime} ,\, E$}
\ArrowLine(140,20)(160,45)
\ArrowLine(180,20)(160,45)
\ArrowArc(160,60)(15,270,90)
\ArrowArcn(160,60)(15,270,90)
\ArrowLine(160,75)(140,100)
\ArrowLine(160,75)(180,100)
\GCirc(160,45){2}{0.2}
\GCirc(160,75){2}{0.2}
\Text(160,0)[]{One Loop: ${\mathcal O}(V^2)$}
\Text(135,20)[r]{$p, \, E$}
\Text(135,100)[r]{$-p^{\prime\prime} ,\, E$}
\Text(185,20)[l]{$-p ,\, E$}
\Text(185,100)[l]{$p^{\prime\prime} ,\, E$}
\Text(140,60)[r]{$p^\prime ,\, E+E^\prime$}
\Text(180,60)[l]{$-p^\prime ,\, E-E^\prime$}
\end{picture}
\caption{Four--electron interaction at tree level and one loop. \label{4elec}}
\end{center}
\end{figure}

Consider a constant four--electron interaction in BCS theory (see
figure \ref{4elec}):
\begin{equation}
V (k_1 , k_2 , k_3 , k_4 ) = {\rm const} = V.
\end{equation}

Concentrate on the one loop term:
\begin{equation}
I = V^2 \int \frac{dE^\prime \, d^2 k^\prime \,
d\ell^\prime}{(2\pi)^4} \; \frac{1}{\left[ (1+i\epsilon)(E+E^\prime) -
v_f (k^\prime) \, \ell^\prime \right]} \frac{1}{\left[
(1+i\epsilon)(E-E^\prime) - v_f (k^\prime) \, \ell^\prime \right]} .
\end{equation}
Here $k$ is the momentum component parallel, and $\ell$ the component
perpendicular to the Fermi surface.

Compute the leading logarithmic divergences of $I$, $I_{\ln}$, and
show that
\begin{equation}
I_{\ln} \sim V^2 \, N\, \ln \left( \frac{E_0}{E}\right) +
{\mathcal O}(V^3),
\end{equation}
where 
\begin{equation}
N \sim \int \frac{d^2 k^\prime}{(2\pi)^3} \; \frac{1}{v_f (k^\prime)}
\sim {\rm density\ of\ states\ at\ Fermi\ energies.}
\end{equation}
Write down a renormalization group flow equation for \[ V(E) =
V - I_{\ln} \] and solve it to show that
\begin{equation}
V(E) \sim \frac{V}{1+N\,V \ln\left( E_0/E \right)} .
\end{equation}
Hence the BCS interactions are marginally relevant if attractive
($ V <0$) and grow stronger as $E\longrightarrow0$.

\subsubsection{BCS vs Phonon--electron pairing.}

Assume a screened Coulomb interaction $V_{\rm c} = {\rm const}$ for
simplicity.  Define its coupling $\mu = N V_{\rm c}$ and
write down a renormalization group flow equation for it.  Repeat the
computation for phonon--electron coupling, and show that it is {\em
not} renormalized.  Discuss the BCS condition for pairing.

\subsection{Landau's Fermi liquid from a renormalization group
viewpoint.}

In a Fermi liquid, the infra red behaviour is governed by a trivial
fixed point \cite{shankar90,shankar94}.  A non Fermi liquid is
characterized and governed by non--trivial infra red fixed points, or
quasi--fixed points (very slow running).

Landau's objective was to study the problem of interacting fermions at
very low temperatures $(T\ll P_{\rm F})$.  He assumed that the system
evolved continuously from the non--interacting limit to the Fermi
liquid theory.  From a renormalization group point of view this is a
trivial (free fermion) fixed point: as one eliminates modes via
$k\leqslant \Lambda$ (the cut--off) one is led to a $1/N_f$ expansion
with $N_f \sim k_{\rm F} / \Lambda$.  The Landau (fixed) point is the
limit $N_f \longrightarrow \infty$.  However, if there are relevant
operators, they can lead to deviations from Landau's fixed point, and
hence from Fermi liquid behaviour.  Theories which are known to
deviate in this way are BCS theory, statistical gauge field theories,
in whose presence Coulombic interactions are modified, and gauge field
theories themselves.

Deviations from Fermi liquid behaviour have been observed in the
normal phase of high--T$_{\rm c}$ cuprates.  It may therefore be
possible to describe these materials using a model governed by a
non--trivial (quasi--) fixed point, {e.g.}\ QED${}_3$ (see sections
\ref{qed3} ff.).

In Fermi liquid theory, the low energy excitations are fermions with a
fermi surface.  The current carrying excitations are quasiparticles,
which have width \[ \Gamma \sim E \left( \frac{E}{E_{\rm F}}\right) \]
where $E$ is some typical energy scale or temperature, and $E_{\rm F}$
is the Fermi energy, characteristic for electrons.  In a
renormalization group sense the Fermi liquid theory has no relevant or
marginal interactions.  There are three main ways in which a model can
differ from normal Fermi liquid behaviour:

\begin{enumerate}
\item[(i)] Marginal interactions: \[ \Gamma \sim E \] which implies
that the electrical resistivity (see section \ref{resistsec}) in the
normal phase of a superconductor is linear in the temperature instead
of quadratic: \[ \rho \sim T .\]
\item[(ii)] BCS instabilities (in the superconducting phase of a
superconductor).
\item[(iii)] Relevant perturbations:  {e.g.}\ deformations of the Fermi
surface.  Note that the issue of naturalness plays a r\^{o}le here:
the deformations of the Fermi surface would have to be fine--tuned
otherwise a small change in (e.g.) the doping would change all the
relevant parameters.
\end{enumerate}

Appealing to experiment \cite{varma89,malinowski97}, the non Fermi
liquid behaviour in the high-T${}_{\rm c}$ shows remarkable stability
up to $T \simeq 600 K$: this excludes (iii) by the fine--tuning
argument \cite{polchinski94}.  With this phenomenology in mind, it has
been argued that the best method is the first, that of marginal
interactions.

\subsection{Concluding remarks.}

The basic features of the renormalization group approach to the Fermi
surface are as follows:

\begin{itemize}
\item Measure momenta from the Fermi surface.
\item Define quasiparticles (``dressed excitations'') with the correct
renormalization group scaling behaviour.
\item Write down the most general low energy effective--theory
interaction terms allowed by the symmetries of the full theory.
\item Compute the scaling behaviour of the interactions as the energy
scale goes to zero, considering quantum corrections (loops).
\item Identify the marginal and relevant interactions:  these are
important for infra red physics.
\item Naturalness requirement for effective field theories: masses
must be constrained to vanish by symmetries or else unnatural fine
tuning is required to cancel masses which will naturally be ${\mathcal
O}(\Lambda)$.  Deformations of the Fermi surface are relevant
parameters, and act like a masses, and therefore must also satisfy the
{\em naturalness} requirement.
\end{itemize}

\section{Lecture II (i):  Non--Trivial Infra Red Structure in QED${}_{\rm
3}$.} \label{qed3}
\setcounter{equation}{0}

As an application of some of the ideas presented in the first
lecture, a large--$N_{\rm f}$, renormalization group improved approach to the
study of U(1) gauge theory in three dimensions is presented.  The
study of systems with a Fermi surface leads to a natural
interpretation of the physics of these gauge theories in terms of a
``flow in theory space,'' associated with the running of the effective
coupling $1/N_{\rm f}$.

Renormalization group improved Dyson--Schwinger computations in three
dimensional U(1) gauge theories have revealed the existence of
non--trivial infra red behaviour as a result of fixed point structure
\cite{aitch+mav:prb,ijra+gac+mkk+dmcn+nm,aitch+mav+mcneill}; see also
reference \cite{kondo97}. This fixed point structure may explain the
non Fermi liquid behaviour of high--T${}_{\rm c}$ superconductors in
their normal phase, or the physics of planar antiferromagnets.

In models with a Fermi surface, renormalization group treatments show
that the effective coupling $1/N_{\rm f}$ is related to the (inverse of the) 
area of
the Fermi surface ({cf.}\ section \ref{quasi}, relation (\ref{Sfarea:N}))
\[
\frac{1}{N_{\rm f}} \longleftrightarrow f\left( \frac{\Lambda}{P_{\rm
F}} \right) .
\]

If the infrared 
fixed point in the theory is trivial, 
then, $1/N_f \rightarrow 0$ and the system is characterized 
by a large Fermi surface.  
If on the other hand, there is a running of $N_f$ such that 
there is no cutoff in the growth of the effective coupling in the infrared, 
then in the deep infra red the Fermi surface will reduce to a point (a
truly relativistic model) as the effective coupling $1/N_{\rm f}$ runs
to infinity.  If however, as in QED${}_3$, there exists a non--trivial
infra red fixed point, then in the deep infra red, the effective
coupling is driven to a finite value and the Fermi surface contracts
to a small pocket.  If the deformations of the Fermi surface are small
then all points on the Fermi surface are equivalent, and one can
linearize about a specific point; this facilitates the use of a
relativistic model, which captures the essential qualitative features.
The relativistic model is the correct one for describing excitations
about the nodes of a $d$--wave superconducting gap.  Such $d$--wave
gaps are known to characterize the physics of high temperature
superconductors \cite{dwave}.  If we apply the slave--fermion
spin--charge separation hypothesis (see section \ref{su2crossu1}) then
the charged excitations about the nodes correspond to Dirac--like
fermions, defining a nodal liquid.  The study of three dimensional
U(1) gauge theories in the normal (chirally symmetric) phase, where
the fermions are massless, serves therefore as a pilot theory for the
quantitative study of the nodal liquid systems and their possible
deviations from Fermi liquid behaviour due to the non--trivial infra
red structure to be discussed below.  Looking at excitations beyond
such $d$--wave nodes is a non--relativistic problem.  As far as normal
phase physics is concerned, however, the arguments at the end of
section \ref{quasi} imply that the dominant interactions are among the
excitations that lie close to one another and therefore the use of
relativistic field theory models may prove to be qualitatively correct
as far as deviations from Fermi liquid behaviour are concerned.  With
this in mind, for the rest of this lecture attention will be
restricted to the study of relativistic gauge field theory models.

A phenomenologically important model for the superconducting phase of
high--T${}_{\rm c}$ cuprates is $\tau_3$--QED${}_3$: a U(1) gauge
theory with the gauge coupling $e$ replaced with
\[
e \longmapsto e\tau_3
\]
so that there are two fermion sectors which couple to the gauge field
with opposite sign:
\begin{equation}
S^{\rm INT}_{\tau_3} = \int d^3 x \; \bar{\Psi} \left(e \tau_3 \asl
\right) \Psi .
\end{equation}
The two fermion sectors correspond to the antiferromagnetic
sub--lattice structure of the underlying condensed matter model
\cite{doreycombo}.  For the normal phase which will be the topic of
this lecture the sub--lattice structure of the underlying model is not
important and from now on it will be ignored.  The resulting theory
therefore will be ordinary QED${}_3$.

It should be noted in passing that the non--trivial fixed point in
QED${}_{\rm 3}$ that we find in our analysis
\cite{aitch+mav:prb,ijra+gac+mkk+dmcn+nm} can be compared with the
ultra violet fixed point in the the three dimensional Thirring model
\cite{debbio97}.  There is the interesting possibility that there
exists a weak--strong coupling (IR $\longleftrightarrow$ UV) duality
which maps between the models.

In the following sections a study of the infra red fixed point
structure of a strongly coupled U(1) gauge theory in three dimensions
is presented.  The non--trivial infra red fixed point structure found
is a {\em non--perturbative} effect, and will be analysed using a
large--$N_{\rm f}$ Dyson--Schwinger treatment.  This method is quite
distinct from conventional Gell--Mann--Low renormalization group
analyses; it is closer in spirit to the Wilsonian effective action
method discussed in the first half of the lectures.

\subsection{Zero temperature analysis of QED${}_3$.} \label{zerotemp}

The three dimensional model considered is a U${}_{\rm S}$(1) gauge theory
of $N_{\rm f}$ 4--component fermion flavours interacting with a statistical
gauge field $a_\mu$.
\begin{equation}
S= \int d^3 x\; \left( \frac{1}{4e^2} F_{\mu\nu}^2 (a) +
\sum_{i=1}^{N_{\rm f}} \bar{\Psi}_i \left( i \dsl + \asl \right)
\Psi_i + {\mathcal L}_{\rm GF} (\xi)\right),
\end{equation}
where the covariant gauge fixing term is given by
\begin{equation} \label{covgauge}
{\mathcal L}_{\rm GF} (\xi) = -\frac{1}{2\xi} \left( \partial_\mu a^\mu
\right)^2 .
\end{equation}
For most of what follows, Landau gauge will be used ($\xi \longrightarrow
0$).

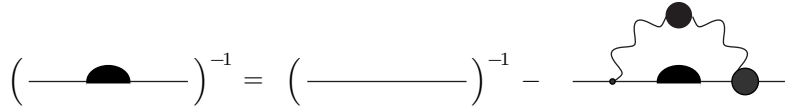
\begin{figure}
\begin{center}
\begin{picture}(300,50)(0,0)
\Text(5,30)[]{$\Big($}
\Line(10,30)(70,30)
\Text(80,31)[]{$\Big)^{-\! 1}$}
\GOval(40,30)(6,8)(0){0}
\CBox(32,23)(48,29){White}{White}
\Text(95,30)[]{$=$}
\Text(110,30)[]{$\Big($}
\Line(115,30)(175,30)
\Text(185,31)[]{$\Big)^{-\! 1}$}
\Text(200,30)[]{$-$}
\Line(215,30)(295,30)
\GOval(255,30)(6,8)(0){0}
\CBox(247,23)(263,29){White}{White}
\PhotonArc(255,30)(25,0,78){-3}{2.5}
\PhotonArc(255.4,30)(25,102,180){-3}{2.5}
\Vertex(255,55){5}
\GCirc(230,30){1}{0.2}
\GCirc(280,30){5}{0.2}
\end{picture}
\caption{Schematic form of the Dyson--Schwinger equation for the full
fermion propagator.  Blobs indicate full non-perturbative
quantities. \label{fermiondseq}}
\end{center}
\end{figure}

In the large--$N_{\rm f}$ limit, a Dyson--Schwinger equation treatment
can be used; the limit is taken in such a way that
\begin{equation}
\alpha \doteq \frac{e^2 N_{\rm f}}{8} = {\rm constant};
\end{equation}
the resulting Dyson--Schwinger equations are shown schematically
in figures \ref{fermiondseq} and \ref{gaugedseq}.  The
Dyson--Schwinger equation for the gauge propagator contains only
graphs of leading order in $1/N_{\rm f}$:  typical graphs not
appearing in the resummation are indicated in figure \ref{not1overN}.

\begin{figure}
\begin{center}
\begin{picture}(400,50)(0,0)
\Photon(0,30)(25,30){3}{2.5}
\Photon(35,30)(60,30){3}{2.5}
\Vertex(30,30){5}
\Text(70,30)[]{$=$}
\Photon(80,30)(130,30){3}{5.5}
\Text(155,30)[]{$+\;\; (-1)$}
\Photon(175,30)(195,30){3}{2.5}
\Photon(215,30)(235,30){3}{2.5}
\CArc(205,30)(10,0,360)
\GCirc(195,30){1}{0.2}
\GCirc(215,30){1}{0.2}
\Text(245,30)[c]{$+$}
\Photon(255,30)(275,30){3}{2.5}
\CArc(285,30)(10,0,360)
\Photon(295,30)(310,30){3}{2}
\CArc(320,30)(10,0,360)
\Photon(330,30)(350,30){3}{2.5}
\Text(360,30)[l]{$+\; \cdots$}
\GCirc(275,30){1}{0.2}
\GCirc(295,30){1}{0.2}
\GCirc(310,30){1}{0.2}
\GCirc(330,30){1}{0.2}
\end{picture}
\caption{Schematic form of the Dyson--Schwinger equation for the gauge
field propagator:  only leading terms in $1/N_{\rm f}$ have been
kept. \label{gaugedseq}}
\end{center}
\end{figure}

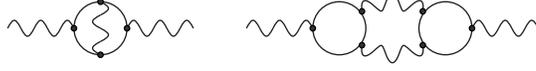
\begin{figure}
\begin{center}
\begin{picture}(200,50) (0,0)
\Photon(0,30)(25,30){3}{2.5}
\CArc(35,30)(10,0,360)
\Photon(45,30)(70,30){3}{2.5}
\GCirc(25,30){1}{0.2}
\GCirc(45,30){1}{0.2}
\Photon(35,40)(35,20){3}{2}
\GCirc(35,40){1}{0.2}
\GCirc(35,20){1}{0.2}
\Photon(90,30)(115,30){3}{2.5}
\CArc(125,30)(10,0,360)
\PhotonArc(145,20)(20,55,125){3}{2.5}
\PhotonArc(145,40)(20,235,305){3}{2.5}
\CArc(165,30)(10,0,360)
\Photon(175,30)(200,30){3}{2.5}
\GCirc(115,30){1}{0.2}
\GCirc(175,30){1}{0.2}
\GCirc(156.47,36.38){1}{0.2}
\GCirc(156.47,23.62){1}{0.2}
\GCirc(133.53,36.38){1}{0.2}
\GCirc(133.53,23.62){1}{0.2}
\end{picture}
\caption{Typical graphs not appearing (to leading order in $1/N_{\rm
f}$) in the resummed Dyson--Schwinger equation for the gauge
propagator (figure \ref{gaugedseq}). \label{not1overN}}
\end{center}
\end{figure}

To proceed, an {\sl ansatz\/} for the fermion propagator is constructed,
and then the Dyson--Schwinger equations are solved for the functions
appearing therein:
\begin{eqnarray} \label{propans}
S_{\rm F}^{-1} (p) &=& -i \big( A(p) \psl + B(p) \big);\nonumber\\
\nonumber\\
A(p) &=& {\rm Wavefunction\ renormalization,}\nonumber\\
B(p) &=& {\rm Gap\ function.}
\end{eqnarray}
The mass gap is given by
\begin{equation}
M\equiv \lim_{p\rightarrow 0} \frac{B(p)}{A(p)},
\end{equation}
and the normal phase is such that $M=0$.  Solving the Dyson--Schwinger
equations leads to coupled integral equations for $A$ and $B$:
\begin{eqnarray} \label{dsgaps}
A(p) &=& 1-\frac{\alpha}{\pi^2 N_{\rm f}}\frac{1}{p^3} \int_0^{\infty}
dk \, \frac{k\, A(k)\, G(p^2,k^2)}{k^2\, A^2 (k) + B^2(k)} {\cal
I}_1(k,p;\alpha); \nonumber \\
B(p)&=&\frac{\alpha}{\pi^2 N_{\rm f}} \frac{1}{p} \int_0^{\infty} dk
\, \frac{k\, B(k)\, G(p^2,k^2)}{k^2 \, A^2 (k) + B^2 (k)} {\cal
I}_2(k,p;\alpha),
\end{eqnarray}
with the integrals given by:
\begin{eqnarray}
{\cal I}_1 (k,p;\alpha) &=& \alpha^2 \ln \left[
\frac{k+p+\alpha}{|k-p|+\alpha} \right] - \alpha \left( k+p -|k-p|
\right) \nonumber \\
&&\quad - \frac{1}{\alpha} |k^2-p^2| \left( k+p - |k-p| \right) + 2kp
\nonumber \\
&&\quad - \frac{1}{\alpha^2} \left( k^2 - p^2 \right) \left\{ \ln
\left[ \frac{k+p+\alpha}{|k-p|+\alpha} \right] - \ln \left[
\frac{k+p}{k-p} \right] \right\} \nonumber ; \\ {\cal I}_2
(k,p;\alpha) &=& 4 \ln \left[ \frac{k+p+\alpha}{|k-p|+\alpha} \right].
\end{eqnarray}
In fact these integrals are heavily damped for $k>\alpha$, and
$\alpha$ can be thought of as an effective (dynamically generated)
ultra violet cut--off.  This dynamical scale arises directly from the
super--renormalizability of QED${}_{\rm 3}$.

The momentum--dependent parts of the full vertex have been written in
terms of the function $G$:
\begin{equation}
\Gamma_\mu (p^2,k^2) = \gamma_\mu \, G(p^2,k^2),
\end{equation}
and this function has to be determined from the Ward--Takahashi identities.

A common {\sl ansatz\/} for the vertex is to write the function $G$ in
terms of the wavefunction renormalization $A$:
\begin{equation}
\Gamma_\mu (p^2,k^2) = \gamma_\mu G(p^2, k^2) = \gamma_\mu A^n (k).
\end{equation}
The value of $n$ is a parameter; the Ward--Takahashi identity can in
principle be used to determine $n$ but there exist kinematical
singularities $1/q^2$ as $q\longrightarrow0$ in the identity.  Here
$n$ will be kept undetermined for the moment, and the issue of the
Ward--Takahashi identity will be ignored.

There are two important features of the above system of integral
equations \cite{kondo+nak92,aitchison94}:
\begin{enumerate}
\item[(i)] An infra red cutoff $\varepsilon$ is required.
\item[(ii)] There exists a critical flavour number $N_{\rm c} = N_{\rm
c} (\varepsilon)$.  As $\varepsilon \longrightarrow 0$, so $N_{\rm c}
\longrightarrow \infty$.
\end{enumerate}

The infra red cut--off can be related to any convenient scale, for
example the temperature, the size of the system, etc.

At low momenta $p\ll\alpha$ a non-zero gap function can be found leading
to a finite dynamical mass which breaks chiral symmetry.  In
particular, for $A=1$ and in Landau gauge the mass is found to be
\begin{equation}
M \sim {\mathcal O}(1) \,\alpha\, e^{- 2\pi/\surd(N_{\rm c}/N_{\rm f}
-1)};
\end{equation}
and for $N_{\rm f} < N_{\rm c}$ there is a chiral symmetry--breaking
dynamical mass.

This analysis is applicable for momenta in the region \[M\ll p \ll
\alpha. \]  In realistic systems ({i.e.}\ field theories describing microscopic
systems with Fermi surfaces) there is another scale in the problem,
namely $P_{\rm F}$. Big Fermi surfaces imply large--$N_{\rm f}$ as modes are
eliminated to pass from lattice to continuum limit(s).

In these condensed matter models the hierarchy of scales is as
follows:
\begin{equation}
M \ll p \ll \alpha \ll P_{\rm F},
\end{equation}
and the ultra violet region in which there is no mass generation is
such that $k\in (\alpha,P_{\rm F})$.  In this regime
\[
N_{\rm f}^{-1} \sim \alpha/P_{\rm F}~ \longrightarrow0.
\]
This is to be contrasted with QED${}_{\rm 3}$ where $\alpha$ behaves
like an ultra violet cut--off, and the kernels of the Dyson--Schwinger
equations die off quickly above this scale: there is no dynamical mass
generation above $\alpha$ in QED${}_3$.

In the low momentum region the kernels of equations (\ref{dsgaps})
can be expanded in powers of $p/\alpha$ and $k/\alpha$ to obtain
\begin{eqnarray} \label{gzeroeq}
A(p) &=& 1 - \frac{g_0}{3} \int_\varepsilon^\alpha dk\; \frac{k\,
A^{n+1} (k)}{k^2 \, A^2 (k) +B^2 (k)} \left[
\left(\frac{k}{p}\right)^3 \theta(p-k) + \theta(k-p)
\right],\nonumber\\
B(p) &=& g_0 \int_\varepsilon^\alpha dk\;\frac{k \, A^{n}(k) \,
B(k)}{k^2\, A^2 (k) + B^2 (k)} \left[ \frac{k}{p} \, \theta(p-k) +
\theta(k-p) \right].\nonumber\\ &&\nonumber\\ g_0 &\equiv&
\frac{8}{\pi^2 N_{\rm f}}
\end{eqnarray}

This system can be studied in the following three momentum regions:

\begin{enumerate}
\item[$p\ll \alpha$] Low energies:  Dynamical mass generation and
chiral symmetry breaking;
\item[$p\sim \alpha$] Intermediate scales:  wavefunction
renormalization in the normal phase;
\item[$p\gg\alpha$] Very high energies:  super--renormalizability at work.
\end{enumerate}

The results of the renormalization group enhanced Dyson--Schwinger
analysis described above lead to the concept of a ``slow running'' of
$N_{\rm f}$ which can be thought of as a flow in theory space
\cite{aitch+mav:prb,ijra+gac+mkk+dmcn+nm}.  As will be shown in
subsequent sections, the dimensionless coupling $g_0$ in equation
(\ref{gzeroeq}) will be renormalized to a running coupling $g_{\rm
R}(p)$.  Specifically, it will be shown that in the infra red region
\( p\ll \alpha \) the following scaling holds:
\begin{equation}
g_{\rm R} \doteq \frac{8}{\pi^2 N_{\rm f} (\alpha,p)} \sim
\left(\frac{p}{\alpha}\right)^{-\gamma} ,
\end{equation}
where $\gamma$ is determined in the subsequent analyses.  In the
context of condensed matter models with a Fermi surface of size
$P_{\rm F}$, this result is to be interpreted as determining the
function $f(\Lambda/P_{\rm F})$ in relation (\ref{Sfarea:N}),
associated with the size of the Fermi surface fundamental cell,
provided that one works in a regime where $P_{\rm F} \sim \alpha$ and
the r\^{o}le of the scale $\Lambda$ is played by the momentum $p$.
Note that this regime should not be confused with the ultra violet
regime $\alpha \ll P_{\rm F}$, which should correspond to the Landau
fixed point.

The presence of the (spontaneous) infra red scale $\varepsilon$
changes the situation at low energies: there exists non--trivial
(quasi--) fixed point structure.  The fixed point is called
``quasi--fixed'' for the infra red cut--off must be removed for the
fixed point to be determined.  The computations show non Fermi--liquid
behaviour, and it is most likely that there is a cross--over between
the two phases ({i.e.}\ not a phase transition).

\subsection{Studies in the $p\gg\alpha$ region.} \label{pggalpha}

\begin{figure}
\begin{center}
\begin{picture}(170,100)(0,0)
\Line(10,10)(10,90)
\Line(50,10)(50,90)
\Photon(10,20)(50,20){3}{2.5}
\Photon(10,30)(50,30){3}{2.5}
\Text(30,55)[]{$\vdots$}
\Photon(10,80)(50,80){3}{2.5}
\Photon(10,70)(50,70){3}{2.5}
\Vertex(10,80){1}
\Vertex(50,80){1}
\Vertex(10,70){1}
\Vertex(50,70){1}
\Vertex(10,20){1}
\Vertex(50,30){1}
\Vertex(10,30){1}
\Vertex(50,20){1}
\Text(75,50)[]{$+$}
\Line(100,10)(100,90)
\Line(140,10)(140,90)
\Photon(100,20)(140,20){3}{2.5}
\Photon(100,30)(140,30){3}{2.5}
\Photon(100,60)(140,80){3}{3}
\CCirc(120,70){2}{White}{White}
\Photon(100,80)(140,60){3}{3}
\Vertex(100,20){1}
\Vertex(140,20){1}
\Vertex(100,30){1}
\Vertex(140,30){1}
\Vertex(100,60){1}
\Vertex(140,60){1}
\Vertex(100,80){1}
\Vertex(140,80){1}
\Text(120,55)[]{$\vdots$}
\Text(155,50)[l]{$+\quad\cdots$}
\end{picture}
\caption{The ladder graphs important in the $p\gg\alpha$ region. \label{ladders}}
\end{center}
\end{figure}
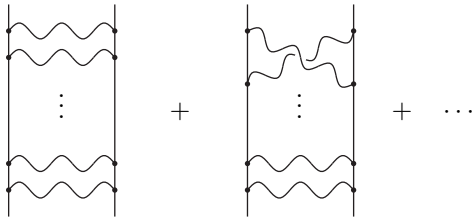

In the very high energy region, the gauge boson polarization tensor
behaves as
\[
\lim_{p\rightarrow\infty}\Pi (p) \sim \frac{\alpha}{8p}
\longrightarrow 0,
\]
and only ladder graphs like those in figure \ref{ladders} are
important \cite{aitch+mav:prb}; this is called the quenched or ladder
approximation.

The wavefunction renormalization in covariant gauge (see equation
(\ref{covgauge}) with arbitrary $\xi$) is given by
\begin{equation}
A(p) = 1 - \frac{1}{3p^2} {\rm Tr} \left[ \psl B(p) \right],
\end{equation}
and the gap function is split into two parts, dependent respectively
on the longitudinal and transverse parts of the vertex
\cite{kondo+nak89}:
\begin{equation}
B(p) = B^{\rm L} (p) + B^{\rm T} (p).
\end{equation}
Now the trace can be evaluated ($x=p^2$), and it is found that the
transverse part vanishes:
\begin{equation}
\frac{1}{3} {\rm Tr} \left[ \psl B^{\rm T} (p) \right] =
\frac{e^2}{8\pi^2} \int_{\varepsilon^2}^{\Lambda^2} dy\; \frac{y^{1/2}
A(y)}{y\, A^2 (y) + B^2(y)} x\, T(x,y),
\end{equation}
for it is a rigorous result that $T(x,y)$ vanishes identically:
\[ T(x,y) = I^0_0 (x,y) - (x-y)^2 I_2^0 (x,y) ;\]
where
\[ I_n^m (x,y) = \int_0^{2\pi} d\theta\; \frac{\sin\theta \,
\cos^m\theta}{\left( x+y-2\surd(xy) \, \cos\theta \right)^2}.\]

Hence only the longitudinal part of the vertex contributes through
$B^{\rm L}$ in quenched QED${}_{\rm 3}$:  again in covariant gauge,
\begin{equation}
A(p) = 1 + \frac{e^2}{8\pi^2} \int_{\varepsilon^2}^{\Lambda^2} dy\; \frac{y^{1/2} \,
A(y)}{y\,A^2(y) + B^2(y)} L(x,y),
\end{equation}
with
\begin{equation}
L(x,y) = \frac{\xi}{x} \left[ (x+y) I_1^0 (x,y) - (x-y)^2 I_2^0 (x,y)
\right] .
\end{equation}
So in Landau gauge ($\xi \longrightarrow 0$) in quenched QED${}_{\rm 3}$
there is no wavefunction renormalization:
\[
A_{\rm \scriptscriptstyle
quenched} (p) =1 ,
\]
and there is no renormalization of the effective flavour number
$N_{\rm f}$.  This corresponds to a trivial fixed point of the
$\beta$--function: in the current model, this occurs at high energies,
and demonstrates that the fermions are {\em asymptotically free} (for
an appropriate choice of the vertex consistent with gauge invariance,
and for $N_{\rm f}\longrightarrow\infty$).

\subsection{Studies in the $p\lesssim \alpha$ regions:  the wavefunction
renormalization.} \label{anomalysec}

In the low and intermediate momentum regions, the wavefunction
renormalization deviates from unity, and becomes important for the
physics of both the normal phase and for dynamical mass generation.

In Landau gauge, the wavefunction renormalization is
\begin{equation}
A(p) \simeq 1 + {\mathcal O}\left( \frac{1}{N_{\rm f}} \right),
\end{equation}
and so is ignored in a resummed $1/N_{\rm f}$ approximation.  With this
prescription it can be shown that there exists a critical flavour
number, $N_{\rm c}$ such that chiral symmetry is dynamically broken
for $N_{\rm f}\leqslant N_{\rm c}$ \cite{appelquist86}.  Note, however, the
criticisms of references \cite{pennington88,atkinson88}.

The precise form of the wavefunction renormalization from the
$1/N_{\rm f}$ resummed graphs is as follows, which shows the critical
exponent (anomalous dimension) indicative of the fixed point
structure:
\begin{equation} \label{anomaly}
A(p) \simeq \left( \frac{p}{\alpha} \right)^{8/3\pi^2 N_{\rm f}} .
\end{equation}
So for $p\sim\alpha$ and in Landau gauge $A(p) \longrightarrow 1$, but
for $p\ll\alpha$ the wavefunction renormalization is relevant for
dynamical mass generation.  $A(p)$ has logarithmic scaling
corrections:  there is no critical flavour number, and chiral symmetry
breaks for all $N_{\rm f}<\infty$.  However this result is not free of
ambiguities, for there remain the  problems of the choice of vertex and of
satisfying the Ward--Takahashi identities.

Regardless of the (non--) existence of a critical flavour number, the
issue of the wavefunction renormalization is crucial, and will be
studied in more detail in the rest of the lectures.

\subsection{The vertex {\sl ansatz\/} and the wavefunction
renormalization.}

In a complete treatment, the vertex {\sl ansatz\/} would be determined
by gauge invariance, through the Ward--Takahashi identity, figure
\ref{wtid}.  This is in general intractable, for the Ward--Takahashi
identity requires the full fermion two--point function for its
solution; in fact, the situation is worse, for there are kinematical
singularities in the identity as $p-k\longrightarrow 0$.  In practice,
then, vertex {\sl ans\"{a}tze\/} are constructed which it is hoped
capture the essential features of the full vertex.

\subsubsection{Simplified treatments.} \label{n=1section}

First, the simplified treatments of references
\cite{kondo+nak92,aitch+mav:prb} are discussed which relate the vertex
to the wavefunction renormalization via a parameter $n$ which is to be
determined:
\begin{equation} \label{webb}
\Gamma_\mu (p^2, k^2) = \gamma_\mu A^n (k) .
\end{equation}
The Pennington--Webb vertex \cite{pennington88}, which does satisfy
the Ward--Takahashi identity as $q = p-k \longrightarrow0$, has $n=1$.

\begin{figure}
\begin{center}
\begin{picture}(300,80)(0,0)
\ArrowLine(10,10)(50,25)
\ArrowLine(50,25)(90,10)
\Vertex(50,25){10}
\Photon(50,35)(50,70){3}{2.5}
\Text(30,30)[]{$p$}
\Text(70,30)[]{$k$}
\Text(130,25)[l]{$(p-k)_\mu \, \Gamma^\mu (k,p) = i S_{\rm F}^{-1} (k) - i
S_{\rm F}^{-1} (p)$}
\end{picture}
\caption{The Ward--Takahashi Identity. \label{wtid}}
\end{center}
\end{figure}

The Dyson--Schwinger equations lead to a definition of an effective
``running'' coupling by the following procedure:

\begin{enumerate}
\item Use a vertex {\sl ansatz}.
\item Apply the bifurcation method (set $B(p)=0$ in the denominator of
the kernel) to the integral equation for $A(p)$.
\item Substitute the solution for $A(p)$ into the equation for $B(p)$.
\item Require running coupling \cite{higashijima84}:
\end{enumerate}
\begin{eqnarray} \label{grenorm}
\frac{e^2}{\alpha} \doteq g_{\rm \scriptscriptstyle R} &\equiv&
\frac{g_0}{A(p,\varepsilon)}; \nonumber\\
g_0 &\equiv& \frac{8}{\pi^2 N_{\rm f}} \Longrightarrow g_{\rm
\scriptscriptstyle R} = \frac{8}{\pi^2 N_{\rm f}(\alpha,p,\varepsilon)} .
\end{eqnarray}
To solve the system of Dyson--Schwinger equations it is essential to
introduce an infra red cut--off (see section \ref{ircutoff}) and use a
Wilsonian renormalization group approach.  As seen in equation
(\ref{grenorm}) this leads to a renormalization of the flavour number
$N_{\rm f}$, and hence a renormalization group flow in theory space.

\subsubsection{More refined treatments.}

Better treatments have been made of the vertex issue for the
wavefunction renormalization \cite{maris96,kondo97}: they involve solving
the Dyson--Schwinger system of equations self--consistently including
the gauge boson polarization, without an infra red cut--off, and with
various vertex {\sl ans\"{a}tze}:
\begin{equation}
\Pi(q) = e^2\, N_{\rm f} \int \frac{d^3 k}{(2\pi)^3} \; \left( 2k^2
-4k\cdot q -\frac{6(k\cdot q)^2}{q^2} \right)
\frac{A(k)}{k^2\,A^2(k)+B^2(k)}\frac{A(p)\,
{\mathcal F}\big(A(p),A(k),A(q)\big)}{p^2\,A^2(p)+B^2(p)},
\end{equation}
where $q=p-k$.

The vertex {\sl ans\"{a}tze\/} are as follows \cite{maris96}
\begin{equation}
\Gamma_\mu = \gamma_\mu \, {\mathcal F} \big( A(p),A(k),A(p-k)\big):
\end{equation}
\begin{eqnarray} \label{fiveans}
&1.& {\mathrm Bare\ vertex:} \ {\mathcal F} =1. \nonumber\\ 
&2.& {\mathcal F}  = \frac{1}{2} \left[ A(p) + A(k) \right] \nonumber\\
&3.& {\mathcal F}  = \frac{A(p)\,A(k)}{A(p-k)} \nonumber\\
&4.& {\mathcal F}  = \frac{1}{4} \left[ A(p) + A(k) \right]^2 \nonumber\\
&5.& {\mathcal F}  = A(p)\,A(k).
\end{eqnarray}
The first and second of these are the only physically motivated vertex
{\sl ans\"{a}tze}; the last is motivated by computational simplicity,
and the others are included to determine how much the vertex choice
affects the results \cite{maris96}.  The third {\sl ansatz\/} yields \[ \Pi(q) \sim
\frac{\alpha q}{A(q)} \] like the $n=1$ {\sl ansatz\/} described
earlier.

\begin{figure}
\begin{center}
\includegraphics[height=0.25\textheight,viewport=5 400 430 590,clip]{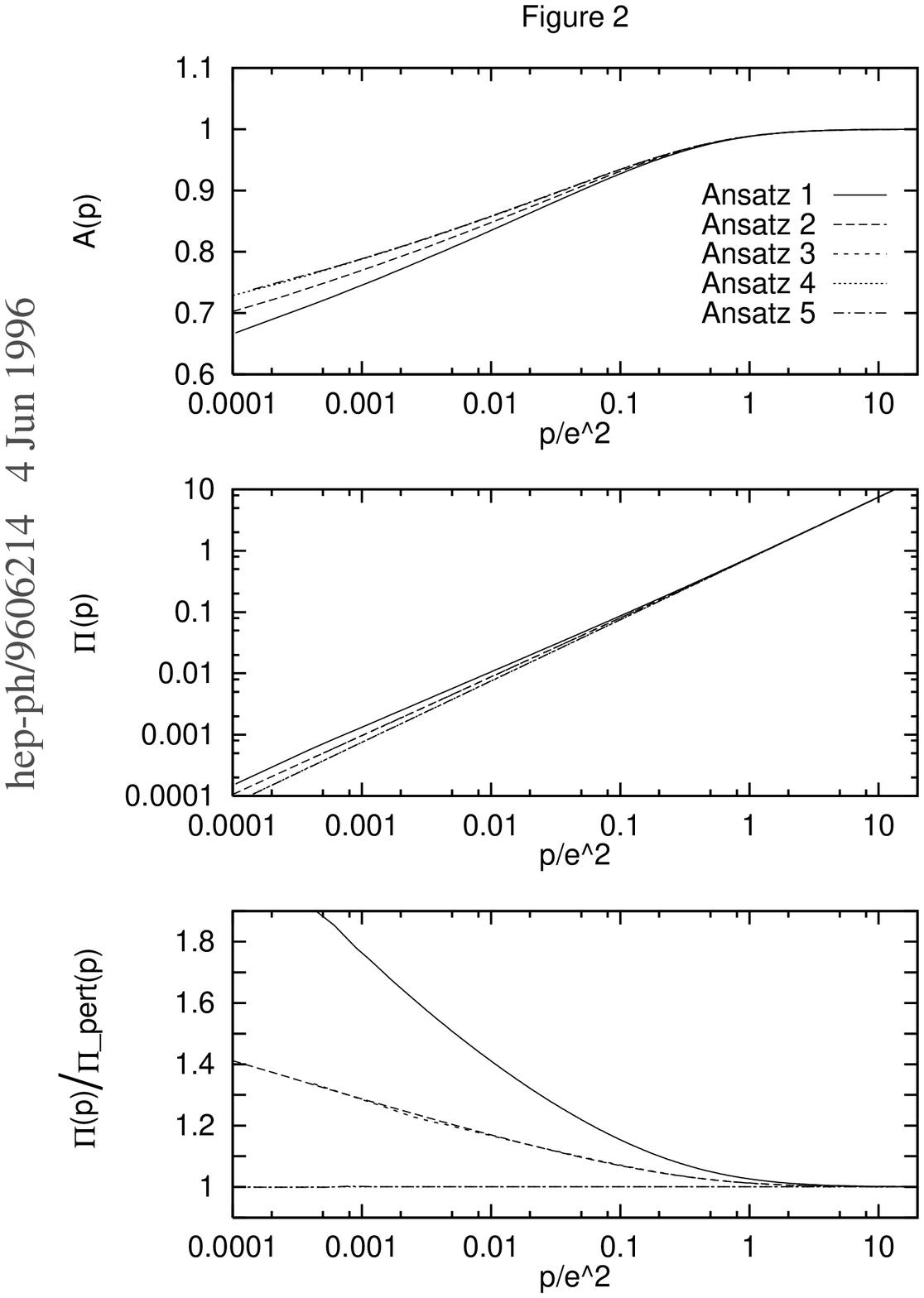}
\caption{The wavefunction renormalization as a function of the
momentum for the five {\sl ans\"{a}tze\/} (\ref{fiveans}). {\sl (After
Maris, \cite{maris96}.)}\label{ansatze}}
\end{center}
\end{figure}

The results of this analysis of the vertex choice are unfortunately
inconclusive (figure \ref{ansatze}), though there is suggestion of the existence of
non--trivial infra red structure, comparable with the critical
behaviour described earlier (equation (\ref{anomaly})).  The behaviour
shown in figure \ref{ansatze} is modified in the presence of an infra
red cut--off, as will be described in the next section.

\subsection{The infra red cut--off.} \label{ircutoff}

There are two main methods for introducing an infra red cut--off to
the system of equations (\ref{dsgaps}):

\begin{enumerate}
\item Wave function renormalization with a momentum space infra red
cut--off $\varepsilon$ directly as a lower limit of integration.
Using the bifurcation method:
\begin{equation}
A(p, \varepsilon) = 1 - \frac{\alpha}{\pi^2 N_{\rm f}} \frac{1}{p^3}
\int_{\varepsilon}^\infty dk \; \frac{{\mathcal I}(p,k)}{k};
\end{equation}
practically speaking, the upper limit of integration is $\alpha$ for
the integrand is heavily damped above this scale.  The variable infra
red scale should be compared with the variable ultra violet scale in a
Wilsonian renormalization group approach.
\item Wavefunction renormalization with covariant infra red cut--off
$\delta$.  Again using the bifurcation method:
\begin{equation}
A(p,\delta) = 1 - \frac{\alpha}{\pi^2 N_{\rm f}} \frac{1}{p^3} \int_0^\infty
dk\; \frac{k\, {\mathcal I}(p,k)}{k^2+\delta^2},
\end{equation}
where again the upper limit is effectively $\alpha$.  This covariant
cut--off is similar to the contribution to the plasmon mass in
finite temperature condensed matter models \cite{aitchison+zuk}.
\end{enumerate}

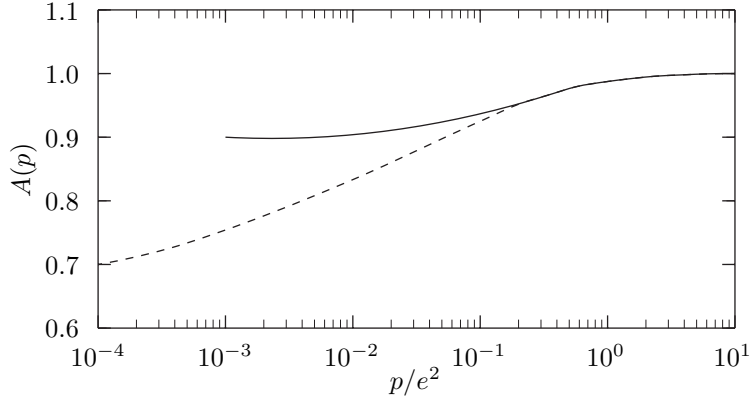
\begin{figure}
\begin{center}
\begin{picture}(260,150)(0,0)
\LinAxis(20,20)(20,140)(5,1,-4,0,0.5)
\LinAxis(260,20)(260,140)(5,1,4,0,0.5)
\LogAxis(20,20)(260,20)(5,4,0,0.5)
\LogAxis(20,140)(260,140)(5,-4,0,0.5)
\Text(0,20)[l]{0.6}
\Text(0,44)[l]{0.7}
\Text(0,68)[l]{0.8}
\Text(0,92)[l]{0.9}
\Text(0,116)[l]{1.0}
\Text(0,140)[l]{1.1}
\Text(20,10)[]{$10^{-4}$}
\Text(68,10)[]{$10^{-3}$}
\Text(116,10)[]{$10^{-2}$}
\Text(164,10)[]{$10^{-1}$}
\Text(212,10)[]{$10^{0}$}
\Text(260,10)[]{$10^{1}$}
\Curve{(68,92)(180,105)(190,108)(200,111)(212,113)(220,114)(230,115)(240,115.5)(250,115.9)(260,116)}
\DashCurve{(20,44)(50,51)(68,57)(116,76)(164,98)(180,105)(190,108)(200,111)(212,113)(220,114)(230,115)(240,115.5)(250,115.9)(260,116)}{3}
\Text(140,0)[]{$p/e^2$}
\rText(-10,80)[][l]{$A(p)$}
\end{picture}
\caption{Schematic representation of the wavefunction renormalization
as a function of the momentum in the presence of an infra red cut--off
$\delta=0.1$ (continuous curve).  The dashed curve indicates the
behaviour with no infra red cut--off from figure
\ref{ansatze}.\label{ircutofffig}}
\end{center}
\end{figure}

The wavefunction renormalization as a function of momentum is shown
schematically in figure \ref{ircutofffig}.  This should be compared
with the dependence without infra red cut--off, figure \ref{ansatze}.

When these infra red cut--offs are removed, neither method gives firm
results for the infra red fixed point.  As $\varepsilon\longrightarrow
0$ the truncated vertex {\sl ansatz\/} plays an important r\^{o}le and the
fixed point cannot be identified.  The removal of $\delta$ is not
smooth; this should be compared with the discontinuities as
$T\longrightarrow 0$ in the plasmon mass model mentioned above
\cite{aitchison+zuk}, and again, the fixed point cannot be identified.

\subsubsection{Graphical results.}

The following four figures are from reference
\cite{aitch+mav+mcneill}; results are presented for $\alpha=1$ and
$N_{\rm f}=5$.

\begin{figure}
\begin{center}
\includegraphics[height=0.2\textheight,viewport=-5 180 450 450,clip]{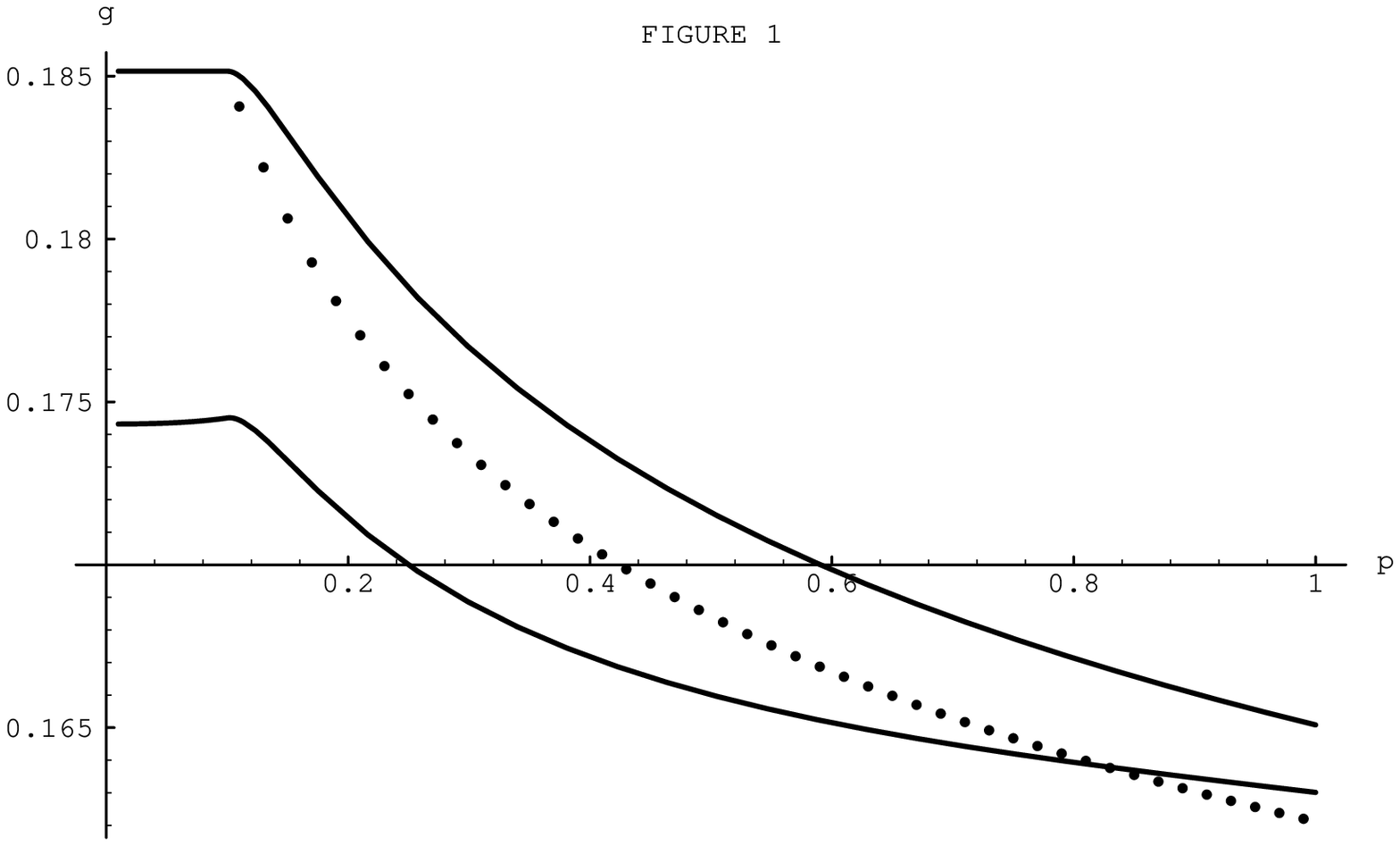}
\caption{The renormalized coupling (ordinate) in the normal phase as a
function of momentum (abscissa) in the presence of a momentum space
infra red cut--off $\varepsilon$ for different computations of the
renormalized coupling.  Here $\varepsilon=0.1$, $\alpha=1$ and $N_{\rm
f}=5$.\label{prbfig1}}
\end{center}
\end{figure}

In figure \ref{prbfig1} the renormalized coupling (\ref{grenorm}) is
plotted as a function of the momentum in the presence of a momentum
space infra red cut--off $\varepsilon$: the upper continuous line is
the analytic result \cite{aitch+mav+mcneill}
\begin{equation} 
g_R = \left\{ \begin{array}{lc} {g_0}/\left({1 - \frac{g_0}{9} +
\frac{g_0}{9}\left( \frac{p}{\alpha}\right)^3 + \frac{g_0}{3} \ln
\left(\frac{p}{\alpha}\right) }\right) \quad & \varepsilon<p<\alpha \\
g_0 / \left( 1+ \frac{g_0}{3} \ln\left( \frac{\varepsilon}{\alpha}
\right) \right) &0<p<\varepsilon \end{array} \right. .
\end{equation}
The dotted curve is a crude approximation \cite{kondo+nak92}:
\begin{equation} 
g_R = \left\{ \begin{array}{lc} g_0 / \left( 1 + \frac{g_0}{3} \ln
\left( \frac{p}{\alpha} \right) \right) \quad & \varepsilon<p<\alpha
\\ g_0 / \left( 1 + \frac{g_0}{3} \ln \left(
\frac{\varepsilon}{\alpha} \right) \right) & 0<p<\varepsilon
\end{array} \right. .
\end{equation}
Finally, the lower continuous curve is the exact (numerical) solution
of equation (\ref{grenorm}).

\begin{figure}
\begin{center}
\includegraphics[height=0.2\textheight,viewport=-5 180 450 450, clip]{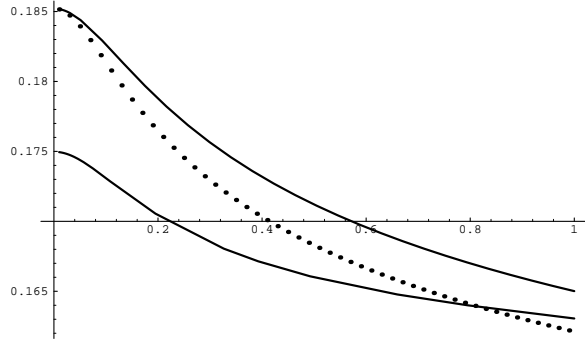}
\caption{The renormalized coupling in the normal phase as a function
of momentum in the presence of a covariant cut--off $\delta$ for
different computations of the renormalized coupling (\ref{grenorm}).
Here $\delta=0.1$, $\alpha=1$ and $N_{\rm f}=5$.\label{prbfig2}}
\end{center}
\end{figure}

In figure \ref{prbfig2} the same results
are presented but for a covariant type cut--off $\delta$.  The upper
continuous curve corresponds to the analytic solution
\begin{equation}
g_R = {g_0}/\left( 1 - \frac{g_0}{9} +
\frac{g_0}{3p^2}\delta^2 - \frac{g_0}{3p^3}\delta^3 \tan^{-1} \left(
\frac{p}{\delta} \right) + \frac{g_0}{6} \ln \left(
\frac{p^2+\delta^2}{\alpha^2+\delta^2}\right) \right),
\end{equation}
while the dotted curve corresponds to the crude approximation
\begin{equation}
g_R = g_0 / \left( 1+ \frac{g_0}{6} \ln \left(
\frac{p^2+\delta^2}{\alpha^2+\delta^2}\right) \right).
\end{equation}
Again, the lower continuous line is the numerical solution of equation
(\ref{grenorm}).

\begin{figure}
\begin{center}
\includegraphics[height=0.2\textheight,viewport=-5 180 450 450,clip]{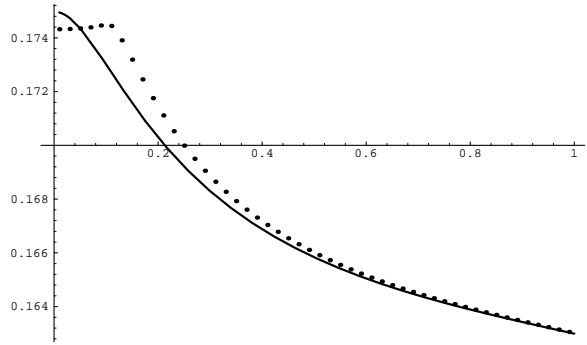}
\caption{The renormalized coupling as a function of momentum for
different types of infra red cut--off.  The dotted and continuous
curves respectively show $g_R (p,\varepsilon=0.1)$ and $g_R (p,
\delta=0.1)$.  Again, $\alpha=1$ and $N_{\rm f}=5$.\label{prbfig3}}
\end{center}
\end{figure}

\begin{figure}
\begin{center}
\includegraphics[height=0.2\textheight,viewport=-5 180 450 450,clip]{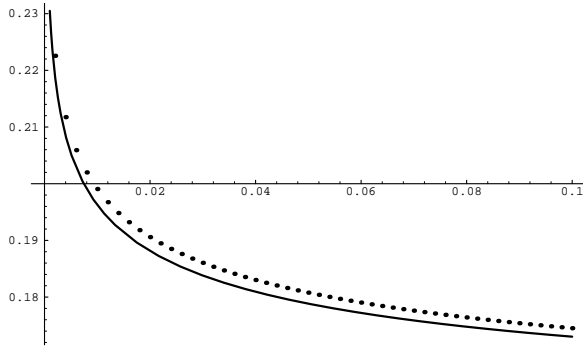}
\caption{The renormalized coupling as a function of the infra red
cut--off.  The dotted and continuous curves respectively show $g_R
(p\sim \varepsilon,\varepsilon)$ and $g_R (p\sim \delta,\delta)$.
Clearly the cut--offs cannot be removed smoothly.\label{prbfig4}}
\end{center}
\end{figure}

A comparative study of the two types of infra red cut--off is shown in
figures \ref{prbfig3} and \ref{prbfig4}.  In the first is a direct
comparison of the running of the coupling with $p$ and the two infra
red cut--offs (the dotted line is for the momentum space cut--off
$\varepsilon = 0.1$).  The second is an indication of how the coupling
behaves as the cut--off is removed in the two schemes: the coupling is
shown as a function of the cut--off, $g_R (p=x,x)$ is plotted against
$x$, where $x \in \{\varepsilon,\delta\}$ (the dotted curve is for
$\varepsilon$).  It is clear that neither cut--off can be removed
smoothly.

\subsubsection{Remark:  generic Wilsonian renormalization approach
to U(1) gauge theory.} \label{t3qed}

\begin{figure}
\begin{center}
\begin{picture}(130,60)(0,0)
\CArc(30,30)(20,0,360)
\CArc(100,30)(20,0,360)
\Photon(80,30)(120,30){3}{3.5}
\GCirc(80,30){1}{0.2}
\GCirc(120,30){1}{0.2}
\LongArrowArcn(30,30)(10,330,50)
\LongArrowArcn(100,30)(15,150,30)
\LongArrowArcn(100,30)(15,330,210)
\end{picture}
\caption{The vacuum graphs in the Wilsonian effective action. \label{wilsonvac}}
\end{center}
\end{figure}
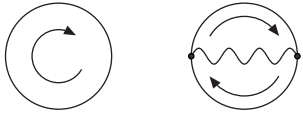

The loops in the Wilsonian effective action (see figure
\ref{wilsonvac}) contain circulating momenta which are cut off at
$M_0$ from above and at $\mu$ from below.  The ultra violet scale
$M_0$ is held fixed and $\mu$ is varied in the infra red.  For graphs
with external lines, terms with external momenta $p<\mu$ are dropped.

If there exist massless states in the theory then it is tricky to
determine the infra red behaviour in this approach (c.f. difference
between Wilsonian and one--particle irreducible effective actions).  The
Wilsonian action is finite but there is running $\sim \ln (p/M_0)$.
In the presence of an infra red cut--off there is extra running in the
infra red with physical consequences.  Such gapless fermionic
excitations exist in the normal phase of (high--T${}_{\rm c}$) superconductors.

This behaviour will persist in the Dyson--Schwinger equations obtained
from the effective action.  This approach can be applied to
investigate infra red (quasi--) fixed point structure and deviations
from Fermi liquid behaviour in theories with fermions.

\subsection{Composite operator effective potential approach.}\label{effpotsec}

The computations presented so far have all been for a $1/N_{\rm f}$
resummed Dyson--Schwinger equation technique, substituting {\sl
ans\"{a}tze\/} for the vertex and fermion propagator.  It is natural to
ask whether the infra red structure found is simply an artifact of the
approximations and the method.  The composite operator effective
potential approach yields a partial answer to this question
\cite{adrian98}.  This approach can be used to investigate the
generality of the fermion {\sl ansatz}, at least to leading order in
$1/N_{\rm f}$.  Whether the structure is a direct result of the
$1/N_{\rm f}$ truncation is discussed in section \ref{beyond1/N}.

The system of Dyson--Schwinger equations for a theory can in principle
be derived from a composite operator effective action
\cite{cornwall74} via a well--defined minimization technique.  If the
{\sl ans\"{a}tze\/} were put straight into the effective action the same
minimization technique could be used to derive the integral
equations for the functions appearing therein.  Experience in other
work \cite{amelino94} has shown that overly restrictive {\sl
ans\"{a}tze} lead to differences between Dyson--Schwinger type
approaches and the results of composite operator effective action
computations.

The composite operator effective action is a generalization of the
conventional (quantum) effective action: it is dependent not only on
possible vacuum expectation values of the quantum fields (as with the
conventional effective action) but also possible vacuum expectation
values of composite operators built from the fields.  The simplest
composite operator effective action is dependent on the expectation
value of the two--point function: for QED${}_3$ we are interested in
probing the propagator {\sl ansatz\/} so this will be sufficient.
This simplest composite operator effective action is the double
Legendre transform of the generating functional for connected Green's
functions.  The composite operator effective potential $V$ is then
defined directly from the composite operator effective action $\Gamma$
by removing a factor of the space--time volume:
\begin{equation}
V \int d^D x\; = -\Gamma.
\end{equation}

For $N_{\rm f}$--flavour QED${}_3$ the series expansion for the
composite operator effective potential in momentum space is
\begin{eqnarray} \label{effpot}
V[S_{\rm F},D] &=&iN_{\rm F} \int \frac{d^3 p}{(2\pi)^3}\; {\rm
tr}\, \left\{ \ln\left[ S_{\rm F}^{-1}(p) \, S_0 (p)\right] + S_0^{-1} (p) \,
S_{\rm F}(p) -1 \right\} \nonumber\\
&-& \frac{i}{2} \int \frac{d^3 p}{(2\pi)^3}\; {\rm
tr}\, \left\{ \ln\left[ D^{-1}(p) \, D_0 (p) \right] + D_0^{-1} (p)\,
D(p) -1 \right\} \nonumber\\
&+& V_2 [S_{\rm F}, D] .
\end{eqnarray}

The functions $S_{\rm F}$ and $D$ are candidate full non--perturbative
two--point functions for the fermions and gauge field respectively,
and the subscripts ${}_0$ denote their bare counterparts.  To determine
the actual (physical) non--perturbative two--point functions for the
theory, the effective potential must be minimized with respect to
(functional) variations in $S_{\rm F}$ and $D$.  Here $-V_2$ is the
sum of all two--particle irreducible vacuum graphs with propagators
set equal to $S_{\rm F}$ and $D$ and with bare (undressed) vertices.
In order to consider more general vertices one would have to consider
a ``trilocal'' effective action (three Legendre transforms) and
require that this new object be stationary with respect to variations
in the vertex {\sl ansatz\/} \cite{adrian98}.

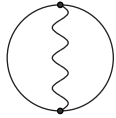
\begin{figure}
\begin{center}
\begin{picture}(60,60)(0,0)
\CArc(30,30)(20,0,360)
\Photon(30,50)(30,10){3}{3.5}
\GCirc(30,50){1}{0.2}
\GCirc(30,10){1}{0.2}
\end{picture}
\caption{Graph contributing to $V_2$ up to two--loop level. \label{2pi}}
\end{center}
\end{figure}

Truncating the series for $V_2$ at two--loop level, figure \ref{2pi},
is sufficient to compare with the Dyson--Schwinger method for
QED${}_3$.  Variation of equation (\ref{effpot}) with respect to
$S_{\rm F}$ and $D$ yields the Dyson--Schwinger equations for
QED${}_3$, which upon substituting the {\sl ansatz\/} (\ref{propans})
gives the equations (\ref{dsgaps}) of section \ref{zerotemp}.  It has
been shown \cite{adrian98} that if one substitutes the {\sl ansatz}
(\ref{propans}) directly into the composite operator effective
potential (\ref{effpot}) then requiring it to be stationary with
respect to (functional) variations in the wavefunction renormalization
$A$ and the gap function $B$ yields {\em the same} integral equations
as the first approach.

Since the composite operator effective potential yields the same
integral equations (\ref{dsgaps}) whether the {\sl ans\"{a}tze\/} are
put in before or after the functional variation demonstrates
explicitly that it is equivalent to the Dyson--Schwinger method for
QED${}_3$.  The consistency of the two methods is a strong indication
that the {\sl ans\"{a}tze\/} are sufficiently consistent and general.

\subsection{The physical consequences of ``slow running.''}

The slow running of the coupling constants (relative to na\"{\i}ve
expectations) enhances the chiral condensate
\begin{equation}
\langle \bar{\Psi} \, \Psi \rangle \sim \int^\Lambda dp\; p B(p).
\end{equation}

If the running of the coupling is slowed down then there is an
enhancement of $B(p)$ at low momenta and therefore of the chiral
condensate.  This is similar to the situation in ``walking
technicolour'' models \cite{appelquist97} where the introduction of
extra fermionic degrees of freedom due to a specific choice of an
enlarged gauge group slows down the running of the pertinent gauge
coupling relative to the model with the original gauge group.

When the field theory is applied to superconductivity models the
enhanced condensate might affect the measurable parameter ${\rm
Gap}/T_{\rm c}$ and the size of the coherence length in magnetic
superconductors.  Also, in the normal phase, where chiral symmetry is
unbroken, the critical behaviour in the wavefunction renormalization
leads to deviations from Fermi liquid behaviour
\cite{ijra+gac+mkk+dmcn+nm}, and also to anomalous electric transport
properties, {e.g.}\ the resistivity linear in $T$ (the temperature) with
$\ln T$ scaling corrections, which behaviour is stable over a large
range of temperatures (see section \ref{last}).  The critical
behaviour in the wavefunction renormalization also modifies the normal
critical behaviour ({i.e.}\ the critical exponents) in these models.

\subsection{Exercises.}

\subsubsection{The Dyson--Schwinger equations.}

Derive the equations (\ref{dsgaps}) from the Dyson--Schwinger
equations for QED${}_3$ to leading order in $1/N_{\rm f}$ (figures
\ref{fermiondseq} and \ref{gaugedseq}).

\subsubsection{Dyson--Schwinger vs composite operator effective
potential.}

Show that the equations (\ref{dsgaps}) can also be obtained from the
composite operator effective potential, equation (\ref{effpot}).

\subsection{Concluding remarks.}

\begin{itemize}
\item Renormalization group improved Dyson--Schwinger analysis reveals
non--trivial infra red fixed point structure in three dimensional U(1)
gauge theory.  This structure in the low energy limit means that,
although QED${}_3$ is super--renormalizable, the degrees of freedom
relevant at scales of order the ultra violet cut--off are still
applicable the in effective theory for the deep infra red.
\item The infra red structure leads to dynamical mass generation and
critical behaviour in the normal (chirally symmetric) phase; these are
associated with a ``slow running'' of the effective coupling.
\item Critical behaviour for the normal phase of the model affects
the wavefunction renormalization
\[
A(p) \sim \left( \frac{p}{\alpha} \right)^{8/3\pi^2 N_{\rm f}}.
\]
\item The computation of this critical behaviour has been discussed in
connexion with a number of vertex {\sl ans\"{a}tze\/} and with different
infra red cut--offs.
\item Slow running of the coupling has physical consequences, such as
in the enhancement of the chiral condensate; some of the physical
consequences of the critical behaviour of the wavefunction
renormalization will be discussed in the last lecture, section
\ref{last}.
\end{itemize}

\section{Lecture II (ii):  Results Beyond Leading Order in $1/N_{\rm
f}$.} \label{beyond1/N}
\setcounter{equation}{0}

A natural question which arises is whether the non--trivial structure
observed in the preceding sections is simply an artifact of the
truncations (to leading order in $1/N_{\rm f}$) in the
Dyson--Schwinger equations.  A partial answer to this question
concerning the fermion {\sl ansatz\/} has already been discussed
(section \ref{effpotsec}); some further attempts to answer this
question and probe the physics beyond leading order in $1/N_{\rm f}$
are presented in what follows.

\subsection{Non--local gauges.}

A treatment of the model beyond $1/N_{\rm f}$ has been performed
recently \cite{kondo97} in which a non--local gauge parameter is
introduced which fixes the wavefunction renormalization $A$ to be
exactly unity.  This approach confirms the earlier results of
references \cite{aitch+mav:prb,ijra+gac+mkk+dmcn+nm} for the existence
of slow running of the effective coupling in the infra red.

The non--local gauge is defined from the gauge fixing term:
\begin{equation}
{\mathcal L}_{\rm GF} = -\frac{1}{2} F\big( a(x) \big) \int d^3 y\;
\frac{1}{\xi(x-y)} F\big( a(y) \big),
\end{equation}
where the function $F$ is given by \[F(a) = \partial_\mu \,a^\mu . \]
The non--local gauge function $\xi$ is then chosen to ensure
$A\equiv1$ exactly.  The results are subsequently transformed back to
Landau gauge using the (inverse) Landau--Khalatnikov transformation
\cite{landau56}.  Both $\varepsilon$ and the covariant $\delta$ type
infra red cut--offs can be used in the computations and the results
compared.

It is found that it is possible to define a running coupling
({cf.}\ equation (\ref{grenorm}))
\begin{equation}
\frac{g_{\rm \scriptscriptstyle R}(t)}{g_{\rm \scriptscriptstyle
R}(0)} \sim 1 + \frac{\tilde{\xi}(k^2)}{2};
\end{equation}
\[ t = \ln \left( \frac{k}{\mu} \right) ;\]
where $\mu$ is some momentum scale and $\tilde{\xi}(k^2)$ is related
to the Fourier transform of the non--local gauge parameter; see
equation (\ref{nlgaugeparam}).

The momentum space cut--off $\varepsilon$ can be removed but when the
covariant type of cut--off is used, there exist discontinuities as
$\delta\longrightarrow0$ ({cf.}\ Landau damping processes at finite
temperature mentioned in section \ref{ircutoff} and in references
\cite{aitchison+zuk,aitchison95}).  In both cases, the infra red fixed
point structure and slow running of the coupling found in previous
analyses \cite{aitch+mav:prb,ijra+gac+mkk+dmcn+nm} is confirmed.

The running of the coupling with the covariant infra red cut--off
$\delta$ is described by the $\beta$--function:
\begin{equation}
\beta(\tilde{\xi}) = -\frac{d}{d t}
\tilde{\xi}(k^2) \sim 4 + \frac{2k^2 +3\alpha k + 4\delta^2}{k^2 +
\alpha k + \delta^2} \left( \tilde{\xi} (k^2) -2 \right) .
\end{equation}

If $\delta = 0$
\begin{equation}
\beta(\tilde{\xi}) = 3 \tilde{\xi}(k^2) -2 \quad : \quad \tilde{\xi}
\stackrel{k\rightarrow0}{\longrightarrow} \frac{2}{3}.
\end{equation}
Compare with $\delta\neq 0$
\begin{equation}
\beta_\delta(\tilde{\xi}) = 4\tilde{\xi}(k^2) - 4 \quad : \quad
\tilde{\xi} \stackrel{k\rightarrow 0}{\longrightarrow} 1;
\end{equation}
so $\delta$ cannot be removed smoothly.

\subsection{Infra red critical exponents beyond $1/N_{\rm f}$ from the
non--local gauge.}

The suggested form of the wavefunction renormalization in a $1/N_{\rm f}$
treatment in Landau gauge is \cite{pennington88} (see section \ref{anomalysec}):
\begin{equation}
A(p) \sim \left( \frac{p}{\alpha} \right)^{\gamma}; \qquad p\ll\alpha
,
\end{equation}
where the critical exponent is given by
\begin{equation}
\gamma = \frac{8}{3\pi^2 N_{\rm f}}.
\end{equation}

In the work of reference \cite{aitch+mav+mcneill} a non--local gauge
is used to verify this form beyond $1/N_{\rm f}$; the computations are
done in a non--local gauge chosen so that $A=1$ exactly and then the
results are transformed back to Landau gauge using the
Landau--Khalatnikov transformation \cite{landau56}.

In the normal phase (chiral symmetry unbroken) and in Landau gauge the
fermion propagator reads:
\begin{equation}
S_{\rm F} (p) = \left[ A(p) \, \psl \right]^{-1}
\end{equation}
and in the non--local gauge in momentum and configuration space
\begin{eqnarray}
S^{\prime}_{\rm F} (p) &=& \left( \psl \right)^{-1}, \nonumber\\
S^{\prime}_{\rm F} (x) &=& i {\not \! x} \frac{
\Gamma\left( 3/2 \right)}{2\pi^{3/2} \, |x|^3 }.
\end{eqnarray}
Now the wavefunction renormalization can be written in terms of the
non--local gauge parameter:
\begin{eqnarray} \label{eqnforAanddelta}
A^{-1} (p) &=& -i \int d^3 x \; e^{i p\cdot x} \, e^{-\Delta(x)} \,
p\cdot x \, \frac{ \Gamma\left( 3/2 \right)}{2\pi^{3/2} \, |x|^3 }
;\nonumber\\
\Delta(x) &\equiv& e^2 \int \frac{d^3 k}{(2\pi)^3} \left( e^{-ik\cdot
x} -1 \right) \frac{ \tilde{\xi}(k^2) }{k^4 \left( 1- \Pi(k^2)/k^2
\right) } ;
\end{eqnarray}
the function $\tilde{\xi}(k^2)$ is related to the Fourier transform of the
non--local gauge parameter:
\begin{equation} \label{nlgaugeparam}
\tilde{\xi}(k^2) = \xi(k^2) \left( 1 - {\Pi(k^2)}{k^2} \right),
\end{equation}
and $\Pi(k^2)$ is the vacuum polarization.  Since in the non--local
gauge $A\equiv 1$ and the trivial (bare) vertex can be chosen while
maintaining consistency with the Ward--Takahashi identity, an {\em
exact} expression for the vacuum polarization can be computed:
\begin{equation}
\Pi(k^2) = - \alpha |\vec{k}|.
\end{equation}
In the deep infra red $|\vec{k}|\ll \alpha$ the non--local gauge
parameter must satisfy
\begin{equation}
\tilde{\xi}(k^2) = 1 + \frac{1}{(k^2)^2 D_{\rm T} (k^2) } \int_0^{k^2}
dz\; z^2 \frac{d}{dz} D_{\rm T} (z) ;
\end{equation}
with the transverse photon propagator given by \[ D_{\rm T} (z) =
\frac{1}{z+ \alpha \surd z}.\] The integral can be performed, to yield
\cite{aitch+mav+mcneill}
\begin{equation}
\tilde{\xi}(k^2) = 2 - 2 \frac{k^2+\alpha k}{k^2} \left[ 1 -
\frac{2\alpha}{k} + \frac{2\alpha^2}{k^2} \ln \left( 1+ \frac{k}{\alpha}
\right) \right]
\end{equation}
and the logarithm can be expanded in powers of $k/\alpha \ll 1$ with
the result
\begin{equation} \label{star}
\frac{\tilde{\xi}(k^2)}{k^4 \left( 1 - {\Pi(k^2)}/{k^2}
\right)} \simeq \frac{2}{3\alpha}\frac{1}{k^3} - \frac{1}{\alpha^2
k^2} + \frac{6}{5\alpha^3 k} .
\end{equation}
For the evaluation of $\Delta(x)$ the Fourier transform of
(\ref{star}) is required:
\begin{equation}
I(x) \equiv \frac{8}{N_{\rm f}} \int \frac{d^3 k}{(2\pi)^3} e^{-ik\cdot x} \left[
\frac{2}{3k^3} - \frac{1}{\alpha k^2} + \frac{6}{5\alpha^2} \right] ;
\qquad e^2 = \frac{8 \alpha}{N_{\rm f}},
\end{equation}
and (see equation (\ref{eqnforAanddelta}))
\begin{equation} \label{deltaofI}
\Delta(x) = I(x) - I(0).
\end{equation}

The integral needs regularization in the ultra violet and the infra
red, so introduce dimensional regulation $d-3 =\epsilon
\longrightarrow 0^+$ and the (renormalization group) scale $\mu$:
\begin{equation}
I(x) = \frac{24}{5\pi^2 N_{\rm f}} \frac{1}{|\alpha \, x|^2}
-\frac{2}{\pi N_{\rm f}} \frac{1}{|\alpha\, x|} + \frac{4}{3\pi^2
N_{\rm f}} \left( \frac{2}{\epsilon} + 2 \ln \left( \frac{1}{|\mu \,
x|}\right) - \gamma_0 \right),
\end{equation}
where $\gamma_0$ is the Euler--Mascheroni constant.  In order to
compute the difference (\ref{deltaofI}) the divergences in $I(0)$ must
be controlled; for this purpose, $I(0)$ is replaced with
$I(1/\alpha)$ \cite{aitch+mav+mcneill}:
\begin{equation}
\Delta(x) = I(x) - I(1/\alpha) \sim \frac{24}{5\pi^2N_{\rm f}} \left(
\frac{1}{x^2} - \alpha^2 \right) - \frac{2}{\alpha\pi N_{\rm f}}
\left( \frac{1}{|x|} - \alpha\right) + \frac{8}{3\pi^2 N_{\rm f}} \ln
\left| \frac{1}{\alpha\, x}\right|,
\end{equation}
which is independent of the ultra violet renormalization scale $\mu$;
this is a direct consequence of the super--renormalizability of
QED${}_3$:  there should only be running in the infra red.

Now the infra red behaviour of the wavefunction renormalization can be
obtained from equation (\ref{eqnforAanddelta}):
\begin{eqnarray}
A^{-1} (p) &=& -i \int d^3 x\; e^{i p\cdot x} p\cdot x \;
\frac{\Gamma(3/2)}{2\pi^{3/2} \, |x|^3} \, e^{-\Delta(x)} \nonumber\\
&\sim& \exp \left( \frac{24}{5\pi^2 N_{\rm f}} - \frac{2}{\pi N_{\rm
f}} \right) \; \frac{2^\gamma \, \pi^{3/2} \,\gamma \, \Gamma(
\gamma/2)}{4\pi \Gamma(\frac{3-\gamma}{2})} \; \left( \frac{\alpha}{p}
\right)^\gamma \qquad (p\longrightarrow 0);\\
\gamma &=& \frac{8}{3\pi^2 N_{\rm f}} \label{critexpt}
\end{eqnarray}

The critical exponent (\ref{critexpt}) confirms the behaviour found in
Landau gauge.  In the approximate method described above, the
prefactor cannot be determined precisely.  This prefactor is
irrelevant and of order unity, indeed for $N_{\rm f}=5$ it has the
numerical value $1.007$.  In section \ref{improved} a more exact
treatment will be described in which the prefactor is determined to be
exactly unity.  The non--local gauge takes the computation of this
critical exponent beyond leading order in $1/N_{\rm f}$ but confirms
the analyses performed to this order.  A treatment of the gap function
in the non--local gauge yields a critical number of flavours
\cite{aitch+mav+mcneill}
\begin{equation}
N_{\rm c} \simeq 4.32
\end{equation}
which is consistent with $1/N_{\rm f}^2$ corrections to the normal Landau
gauge computations \cite{nash89}.

\subsection{Improved computation of behaviour beyond $1/N_{\rm f}$.} \label{improved}

A more refined treatment using the non--local gauge has been performed
\cite{kondo+mura97} in which a Dyson--Schwinger type equation for the
fermion propagator $S$ is derived in place of the normal equation for
$S^{-1}$.  Consider the equations for the fermion propagator in the
non--local gauge ($S_0$ is the free fermion propagator) and in $D$ dimensions:
\begin{equation}
S_{\rm NLG} = S_0 \qquad : \qquad i\!\!\dsl S_0 = \delta^D (x).
\end{equation}
In Landau gauge,
\begin{eqnarray}\label{landauprop}
S_{\rm L} (x) &=& e^{-\Delta(x)} S_{\rm NLG} (x) = \left[ A(p) \, \psl
\right]^{-1} ,\\
\Delta(x) &=& e^2 \int \frac{d^D k}{(2\pi)^D} \left( e^{i k\cdot x} -1
\right) f(k) , \nonumber\\
f(k) &=& \frac{\xi(k)}{k^4} \nonumber.
\end{eqnarray}

Now, by acting on equation (\ref{landauprop}) with $i\!\!\dsl$, a
Dyson--Schwinger type equation for the fermion propagator in the
normal phase can be derived (using the fact that $\Delta(0) = 0$):
\begin{eqnarray}
i\!\!\dsl S_{\rm L} (x) &=& \delta^D (x) - i e^2 S_{\rm L} \int \frac{d^D
k}{(2\pi)^D} \; \not\! k \, f(k) e^{-ik\cdot x};\nonumber\\
\Longrightarrow A^{-1} (p) &\doteq& Z(p) = 1 + e^2 \int \frac{d^D q}{(2\pi)^D}\;
\frac{q\cdot (q-p)}{q^2} \, f(q-p) \, Z(q).\label{twostar}
\end{eqnarray}

The solutions of equation (\ref{twostar}) are as follows \cite{kondo+mura97}
\begin{eqnarray} \label{L_Deqns}
Z(p) &=& 1 - e^2 \int_\varepsilon^\Lambda dq\; Z(q) \, L
(p,q;D),\nonumber\\
L (p,q;D) &=& \frac{q^{D-3}}{2^{D-1} \pi^{(D+1)/2}
\Gamma(\frac{D-1}{2})} \int_0^\pi d\theta \; \sin^{D-2} \theta \;
\left( pq\cos\theta -q^2 \right) \, f\left( \sqrt{ (p-q)^2} \right) ;
\nonumber\\
f(k) &=& \frac{2}{k^4+\alpha k^3} - \frac{2}{k^4} \left[ 1 -
\frac{2\alpha}{k} + \frac{2\alpha^2}{k^2} \ln \left(1+\frac{k}{\alpha}
\right) \right].
\end{eqnarray}
Note that in three dimensions $L(q,p;3)$ is not a symmetric function
of its arguments.  In the infra red, $k\ll\alpha$ the regularization
scale $\Lambda$ can be taken equal to the effective cut--off $\alpha$,
and the infra red cut--off $\varepsilon$ can be taken to zero.  In
this regime, the equation (\ref{L_Deqns}) can be solved numerically
and analytically using truncated expansions.  In the absence of the
infra--red cut--off, $\varepsilon\longrightarrow 0$,
\begin{equation}
Z(p) = A^{-1} (p) = \left( \frac{p}{\alpha}
\right)^{-8/3\pi^2 N_{\rm f}}
\end{equation}
which is {\em exact\/} and the same as the result obtained in
reference \cite{aitch+mav+mcneill}.  It is an open issue as to whether
these results are dependent on the type of infra red cut--off, {i.e.}\
whether the results are the same if a covariant cut--off is included:
\[ \frac{1}{k^2} \longmapsto \frac{1}{k^2 + \delta^2}.\]

\subsection{An alternative description of running flavour number.} \label{papsec}

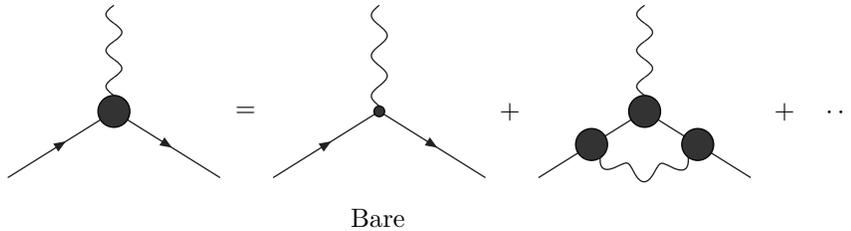
\begin{figure}
\begin{center}
\begin{picture}(350,90)(0,0)
\ArrowLine(10,20)(50,45)
\ArrowLine(50,45)(90,20)
\GCirc(50,45){6}{0.2}
\Photon(50,51)(50,85){3}{2.5}
\Text(100,45)[]{$=$}
\ArrowLine(110,20)(150,45)
\ArrowLine(150,45)(190,20)
\GCirc(150,45){2}{0.2}
\Photon(150,47)(150,85){3}{2.5}
\Text(200,45)[]{$+$}
\Line(210,20)(250,45)
\Line(250,45)(290,20)
\GCirc(250,45){6}{0.2}
\Photon(250,51)(250,85){3}{2.5}
\GCirc(230,32.5){6}{0.2}
\GCirc(270,32.5){6}{0.2}
\PhotonArc(250,45)(23.58,225,314){3}{2.5}
\Text(300,45)[l]{$+\quad \cdots$}
\Text(150,5)[]{Bare}
\end{picture}
\caption{Pinch technique for renormalizing the coupling $G$ from
the amputated vertices. \label{pinch}}
\end{center}
\end{figure}

To conclude this section, a comment on the correctness of the
effective running coupling inferred above from the special
manipulations of the Dyson--Schwinger equations (\ref{dsgaps}) in the
normal phase.

In field theory the actual running coupling constant is usually defined
via the amputated vertex function.  There is work currently underway
\cite{nm+papa} on renormalizing the coupling using a ``pinch
technique'' similar to that used for 4--dimensional QCD
\cite{cornwall91}.  In place of the Dyson--Schwinger equations for the
fermion propagator, the Dyson--Schwinger equation for the amputated
vertex is investigated (figure \ref{pinch}) and the running of the
coupling determined directly.  The computations are performed in a
large--$N_{\rm f}$ framework, with $e^2 N_{\rm f}$ held fixed.  Then
the dimensionless coupling runs, and generates an effective running of
$N_{\rm f}$:
\begin{equation}
\left( \frac{e^2}{\alpha} \right)_{\rm ren} \sim \left(
\frac{1}{N_{\rm f}} \right)_{\rm ren} .
\end{equation}
The differential equation for the vertex function $G(p)$ is
complicated by the presence of infra red cut--offs, and can only be
solved numerically at present.  The advantage of this method lies in
the fact that in determining the vertex directly, the problem of
finding vertex {\sl ans\"{a}tze\/} satisfying the Ward--Takahashi does
not arise.

\subsection{Concluding remarks.}

\begin{itemize}
\item Non--local gauges have been used to probe the infra red physics
of three dimensional U(1) gauge theory beyond leading order in
$1/N_{\rm f}$.
\item The results have been shown to be consistent with those obtained
in conventional $1/N_{\rm f}$ Dyson--Schwinger treatments.  This
supports the idea that the non--trivial infra red structure found is
not an artifact of the $1/N_{\rm f}$ truncation.
\end{itemize}

\section{Lecture II (iii):  Predictions of Gauge Interactions for Finite Temperature
Models.}\label{last}
\setcounter{equation}{0}

\subsection{The resistivity in gauge field theory models.}

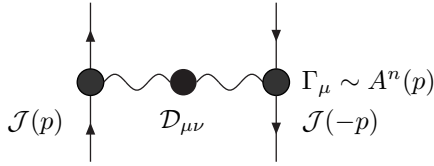
\begin{figure}
\begin{center}
\begin{picture}(110,80)(0,0)
\ArrowLine(20,10)(20,35)
\ArrowLine(20,45)(20,70)
\GCirc(20,40){5}{0.2}
\Photon(25,40)(50,40){3}{1.5}
\Vertex(55,40){5}
\Photon(60,40)(85,40){3}{1.5}
\GCirc(90,40){5}{0.2}
\ArrowLine(90,70)(90,45)
\ArrowLine(90,35)(90,10)
\Text(10,25)[r]{${\mathcal J}(p)$}
\Text(100,25)[l]{${\mathcal J}(-p)$}
\Text(55,25)[]{${\mathcal D}_{\mu\nu}$}
\Text(100,40)[l]{$\Gamma_\mu \sim A^n (p)$}
\end{picture}
\caption{Current--current correlator for the resistivity
computation; the wavefunction renormalization affects the resistivity
through the vertex. \label{current-current}}
\end{center}
\end{figure}

In this final part of the lectures the preceding ideas will be applied
to a determination of some properties of QED${}_3$ at finite
temperature.  Attention will be restricted to the predictions of the
gauge theory model for the electrical resistivity.  The resistivity is
defined as the response of a system to a change in external electric
field (see section \ref{resistsec}).  Formally it is related to the
imaginary part of the electric current--current correlator:
\begin{equation} \label{resisteq}
({\rm Resistivity})^{-1}=({\rm conductivity}) \sim {\mathcal I}m \,
\left. \left\langle {\mathcal J}_\mu^{\Psi} (p) \, {\mathcal
J}_\nu^{\Psi} (-p) \right\rangle \right|_{p=0} \propto {\mathcal I}m
\, \left.\left( A^n (p) \, {\mathcal D}_{\mu\nu}(p) \, A^n (p) \right)
\right|_{p=0},
\end{equation}
where $A(p)$ is the wavefunction renormalization (which appears here
as a result of the vertex {\sl ansatz\/}, see equation (\ref{webb})
and figure \ref{current-current}) and ${\mathcal D}_{\mu\nu}$ is the
full gauge field propagator.  The current is defined by
\[
{\mathcal J}^{\Psi}_\mu = \frac{\delta S^{\rm eff}}{\delta A^\mu_{\rm
ext}},
\]
with $A^\mu_{\rm ext}$ an external electromagnetic potential.  As was
observed in section \ref{n=1section}, gauge invariance selects the
Pennington--Webb vertex with $n=1$.

The gauge field theory model of superconductivity considered above
gives rise to two contributions to the resistivity
\cite{aitch+mav:prb}: the bulk effect of the gauge field and the
logarithmic corrections due to the wavefunction renormalization from
the vertices (figure \ref{current-current}).  We shall review briefly
these results in what follows.

To be completely rigorous in deriving equation (\ref{resisteq}) the
precise behaviour of the effective action $S^{\rm eff}$ in the infra
red is required.  However, as will be discussed in section
\ref{realtimesec} there are non--analyticities due to Landau damping
\cite{aitchison+zuk} which render the infra red limit intractable in
the low temperature limit, $T\longrightarrow 0$.  To circumvent this
problem a physical (heuristic) approach will be used, exploiting the
spin--charge separation (see sections
\ref{su2crossu1}--\ref{vacuasec}) in the framework of Ohm's law.  The
method used is summarized as follows:

\begin{itemize}
\item ``Spin--charge separation:'' (see section \ref{su2crossu1})
split the electronic degrees of freedom into holon (charge) and spinon
(spin) \cite{anderson87} degrees of freedom ($\Psi^{\dagger}$) and
($Z_\alpha$) respectively
\begin{equation}
C_\alpha = \Psi^{\dagger} \, Z_\alpha
\end{equation}
\item Assume Ohm's Law for an external electric field $\vec{E}$:
\begin{equation}
{\mathcal J}^\Psi = (charge) \times v_{\rm F} = \vec{\sigma}\cdot\vec{E},
\end{equation}
where ${\mathcal J}^{\Psi}$ is the current for the holon degrees of
freedom.  Here $v_{\rm F}$ is the Fermi velocity for holons and is a
function of temperature: this temperature dependence arises from
non--trivial thermal vacua (in the relativistic limit of
antiferromagnets); see section \ref{vacuasec}.
\end{itemize}

The results can be summarized as:
\begin{enumerate}
\item The bulk effect of the gauge interactions
(if the wavefunction renormalization $A(p)$ is unity) is to give a
temperature dependence to $v_{\rm F}$ of the form
\begin{equation}
v_{\rm F} \propto T^{-1} \qquad {\rm for\  low\  } T.
\end{equation}
\item For $A(p) \neq 1$
\begin{equation}
\rho \sim \frac{1}{\sigma} \sim T^{1 - {\mathcal O}(1/N_{\rm f})}.
\end{equation}
\end{enumerate}

The non--trivial infra red structure encoded in the wavefunction
renormalization corrects the normal linear behaviour.  As was
discussed above there are gauges in which the wavefunction
renormalization is trivial.  The resistivity $\rho$ is physical, and
should therefore be gauge invariant: in gauges where the wavefunction
renormalization is trivial there are still logarithmic corrections in
the resistivity arising from non--trivial structure in the resulting
vertex functions.

\subsection{The real time formalism and Landau damping.} \label{realtimesec}

The action for massive fermions interacting with a U${}_{\rm S}$(1) statistical
gauge field $a$ and an external electromagnetic field $A^{\rm em}$ in
three dimensions is
\begin{equation} \label{3dfermion-emaction}
S = \int d^3 x \; \bar{\Psi} \left( i \dsl - e {\not \!\! A}^{\rm em} -g_s \!\!\asl
- M \right) \Psi ;
\end{equation}
it is sufficient to study this model with the statistical gauge field
turned off.  The one--loop effective action can be computed:
\begin{eqnarray}
W^{(1)} &=& \frac{ie^2}{2} \int \frac{d^3 p}{(2\pi)^3} \;
\tilde{A}^{\rm em}_\mu (-p) \, \tilde{\Gamma}^{\mu\nu} (p) \,
\tilde{A}^{\rm em}_\nu (p) ;\nonumber\\
\tilde{\Gamma}^{\mu\nu} (p) &=& \int \frac{d^3 k}{(2\pi)^3} \; {\rm
tr}\, \left[ \gamma^\mu \, S_{\rm F} (k+p) \, \gamma^\nu \, S_{\rm F}
(k) \right].
\end{eqnarray}

As an example of non--analyticities near the origin in $p$--space,
consider the odd parity part of $\tilde{\Gamma}$:
\begin{eqnarray}
\tilde{\Gamma}^{\mu\nu}_{\rm odd} (p) &=& -2 F(p) \,
\epsilon^{\mu\nu\lambda}\, p_{\lambda}\nonumber\\
F(p) &=& -i M\int \frac{d^3 k}{(2\pi)^3} \; \frac{1}{\left[ (k+p)^2
-M^2 \right] \left[ k^2 -M^2\right]} .
\end{eqnarray}

Here the imaginary time formalism is used to compute finite
temperature effects:  perform an analytic continuation to Minkowski
space: \[ p_3 \longmapsto -i p_0. \] The result after such a
continuation is (with $\beta$ the inverse temperature)
\begin{eqnarray}
F_\beta (p_0 , \vec{p}) &=& \frac{M}{4} \int \frac{d^2 k}{(2\pi)^2} \;
\left[ \frac{n_{\rm F} (E_k) + n_{\rm F} (E_{k+p})}{p_0 - E_k -
E_{k+p}} - \frac{ n_{\rm F}(E_k) + n_{\rm F} (E_{k+p})}{p_0 +E_k
+E_{k+p}} - \frac{n_{\rm F}(E_k) - n_{\rm F}(E_{k+p})}{p_0-E_k
-E_{k+p}}\right.\nonumber\\
&&\qquad\qquad + \left.\frac{n_{\rm F}(E_k) -n_{\rm F}(E_{k+p})}{p_0+E_k
-E_{k+p}} \right],
\end{eqnarray}
where \[ n_{\rm F} (E) \doteq \tan \frac{\beta E}{2} = -i \left[ 1-
2n_{\rm F}(iE) \right] \] is the Fermi--Dirac distribution.

The denominators of $F$ have discontinuities along the real axis,
which lead to delta functions and contributions to the imaginary parts
of the effective action $F_\beta$ and hence to physical processes,
{e.g.}\ fermion $\longleftrightarrow$ fermion $+$ gauge quantum:
\u{C}erenkov processes or (in many body language) {\em Landau
damping.}
\[
\delta(p_0 + E_k -E_{k+p}) \quad : \quad E_{k+p} = \left[ M^2 +
(\vec{k} + \vec{p})^2 \right]^{1/2} \simeq E_k +
\frac{\vec{p}\cdot\vec{k}}{E_k}
\]
Near the origin of momentum space the $\delta$--function condition
leads to
\begin{equation}
p_0 = \frac{\vec{p}\cdot\vec{k}}{E_k} = \vec{p}\cdot\vec{v_k}
\end{equation}
where $v_k$ is the fermion velocity; and there is a discontinuity in
the space--like region \(p_0 \leqslant |\vec{p}|. \)

There are non--analyticities associated with such processes.  The
argument in qualitative terms follows.  The contribution to the
imaginary part of $F_\beta$ involves angular integrations for both
$\vec{p}$ and $\vec{k}$.  But there exists a cut in the $p_0$ plane
which extends from $-|\vec{p}|$ to $|\vec{p}|$.  So for
$|\vec{p}|\longrightarrow\vec{0}$ (which is a relevant limit for the
computation of the resistivity, equation (\ref{resisteq})) there is
non--analyticity.  It is worth noting in passing that this situation
is analogous to the case of BCS superconductivity for $T<T_{\rm c}$ in
a time dependent Ginzburg--Landau theory.  This has also been verified
by explicit computation \cite{aitchison+zuk}.

The non--analyticities described above result in a non--local effective
action and complicate the situation enormously.  A way out of this
problem is to use approximate closed expressions for low $p$ and then
numerically verify them up to momenta and energies of order of the
fermion mass scale in the particular problem.  In this way it is
possible to construct an effective action that is local for some
region of parameter space.

A more physical way out of this problem is to use a heuristic approach
involving spin--charge separation in conjunction with Ohm's law which
is the subject of the next subsection.

\subsection{Spin--charge separation and resistivity.}

\subsubsection{Resistivity and Ohm's law.} \label{resistsec}

Consider the case of QED${}_3$ in the presence of an external electric
field $\vec{E}$ corresponding to an electromagnetic potential $A^{\rm
em}_\mu$.  Starting with the action (\ref{3dfermion-emaction}) and
integrating out all fields but the electromagnetic:
\begin{eqnarray}
S &=& \int d^3 x\; \bar{\Psi} \left( i \dsl - \asl - e
\not{\!\!A}_{\rm em} \right) \Psi \nonumber\\
S_{\rm eff} &=& \int d^3 x\; A_{\rm em}^\mu (p) \, {\mathcal D}_{\mu\nu} \,
A_{\rm em}^\nu (-p) ; \nonumber\\
{\mathcal D}_{\mu\nu} &=& \left( \delta_{\mu\nu} - \frac{p_\mu
p_\nu}{p^2} \right) \frac{1}{\vec{p}^2 + \Pi}.
\end{eqnarray}

The electric current is given by
\begin{equation}
\frac{\delta S_{\rm eff}}{\delta A_{\rm em}^i} \equiv {\mathcal J}^{\rm
el}_i \propto E_i(\omega).
\end{equation}
In momentum space the electric field $E_i (\omega)$ is given by
$\omega \, A^{\rm em}_i$.  Hence Ohm's law in a gauge with $A^{\rm
em}_0 =0$ gives the conductivity as \cite{ioffe89}
\begin{equation}
\sigma_f = \left.\frac{1}{\vec{p}^2 + \Pi}\right|_{\vec{p}=\vec{0}} .
\end{equation}

\subsubsection{The spin--charge separation formalism.}
\label{su2crossu1}

In the slave fermion spin--charge separation formalism
\cite{spincharge,doreycombo} the electron degrees of freedom are split
into two parts at each lattice site $n$ of the condensed matter model:
\begin{equation} \label{spinchargeeq}
C_\alpha^n = \Psi_n^{\dagger} Z^n_\alpha ,
\end{equation}
where $\Psi$ is a spinless electrically charged fermion (hole) and
$Z_\alpha$, $\alpha \in \{1,2\}$ is a $CP^1$ magnon which carries the
spin part of the degrees of freedom.

There is a constraint of at most one electron per lattice site which
is expressed as
\[
\Psi_n^\dagger \Psi_n + (Z^*_n){}_\alpha (Z_n)_\alpha =1, \qquad
({\mathrm no\ sum\ over\ }n).
\]
In view of this, the currents are related by
\begin{equation}
\vec{{\mathcal J}}^\Psi + \vec{{\mathcal J}}^Z = \vec{0} ,
\end{equation}
where
\begin{equation}
\vec{{\mathcal J}}^\Psi = \bar{\Psi} \, \vec{\gamma}\, \Psi
\qquad\qquad \vec{{\mathcal J}}^Z = Z^* \, \vec{\partial} \, Z .
\end{equation}

The movement of the charge carriers is governed by the transport
velocity, which is that of the $CP^1$ gauge fields, {i.e.}\ in the
relativistic effective field theory.  Here $v_f$ plays the r\^{o}le of
the ``speed of light,'' insofar as it is a limiting velocity for the
low energy excitations.

\begin{figure}
\begin{center}
\begin{picture}(120,50)(0,0)
\CArc(60,30)(20,0,360)
\Photon(0,30)(40,30){3}{3.5}
\Photon(80,30)(120,30){3}{3.5}
\GCirc(80,30){1}{0.2}
\GCirc(40,30){1}{0.2}
\end{picture}
\caption{The fermion vacuum polarization contribution to the gauge
boson two--point function.\label{qedvacpol}}
\end{center}
\end{figure}
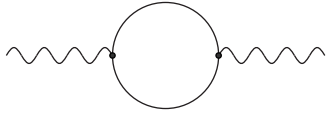

In order to compute the resistivity we use the phenomenological Ohm's law:
\begin{equation} \label{ohmslaweq}
{\mathcal J}^\Psi = {\rm charge} \times v_f =
\vec{\sigma}\cdot\vec{E}.
\end{equation}
As will be shown next the Fermi velocity $v_f$ is a function of
temperature in non--trivial thermal vacua \cite{latorre95}.  A
non--trivial thermal vacuum arises in finite temperature QED${}_3$ due
to fermion vacuum polarization, see figure \ref{qedvacpol}.

\subsubsection{Speed of light in non--trivial vacua.}
\label{vacuasec}

The effective speed of light can be computed \cite{aitch+mav:prb} from
the dispersion relation induced by the fermion vacuum polarization
(see figure \ref{qedvacpol})
\begin{equation}
v_f^{\rm eff} = \frac{\partial E}{\partial Q} ,
\end{equation}
where $E$ is the energy of off--shell gauge bosons ({cf.}\ on--shell
Landau processes as the main contributions to microscopic
resistivity):
\begin{equation}
E^2 \equiv q_0^2 = Q^2 + \Pi(Q, \beta),
\end{equation}
and $\Pi$ is the thermal vacuum polarization (pertaining to the real
part of the effective action).  For on--shell gauge bosons the
denominator of the propagator vanishes.

An approximate {\sl ansatz\/} (found to be a good approximation
numerically) which qualitatively captures the correct behaviour is
\cite{aitchison94}
\begin{eqnarray}
\Pi(Q,\beta) &\sim& \Pi_{\rm L} \simeq \Pi_{\rm T} \simeq \left(
\frac{\alpha p^2}{64} + 4 \, \omega^4_\pi \right); \nonumber\\
p^2 &\equiv& p^2_3 + \vec{p}^2
\end{eqnarray}
with $\omega_\pi$ the plasmon mass.  It is found that this approximate
form is applicable over a large temperature range.  Then
\begin{equation}
v_f \propto \frac{Q}{T^{3/2}}:  \qquad Q\longrightarrow
\varepsilon_{\rm IR} \sim \surd (\alpha/\beta) \propto \surd T .
\end{equation}
Hence the behaviour of the resistivity is easy to infer from equation
(\ref{ohmslaweq}), assuming $T$--independence of the electric field $\vec{E}$,
\begin{equation}
v_f \sim T^{-1} \Longrightarrow \rho \propto T .
\end{equation}
This type of resistivity response in the bulk is characteristic of a
gauge interaction, and strongly suggests that the gauge model
correctly describes the physics in these materials.  Indeed it is very
hard to arrange this behaviour in models with non--gauge interactions.

\subsubsection{Logarithmic corrections to the resistivity.}

The linear--$T$ behaviour demonstrated above has logarithmic
corrections coming from the wavefunction renormalization.  To compute
these it is only necessary to look at the real parts of the effective
action.  The Dyson--Schwinger equation for the wavefunction
renormalization at finite temperature is as follows
\begin{equation}
A(P,\beta) \simeq 1 + \frac{\alpha^2}{16\pi^2 N_{\rm f}} \int_0^\alpha dK \;
I(P,K,\beta) \, \frac{ \tanh\left[ \frac{\beta}{2} \, \surd \left( K^2 +
M^2(K,\beta) \right)\right]}{\surd \left( K^2 + M^2(K,\beta) \right)}.
\end{equation}
Here the mass function $M(K,\beta)$ is simply the pole mass, given by
the ratio of the gap function to the wavefunction renormalization, and
$P,K$ denote spatial momentum components.

To continue it is necessary to make an {\sl ansatz\/} for the kernel:
replace the component $\Pi_{00}$ with a constant $\Delta \sim
{\mathcal O}(\alpha^2)$ \cite{aitchison94}.  With this {\sl ansatz}
\begin{equation}
I = -\frac{2\pi}{P^2} \left\{ 1 - \frac{|P^2-K^2|}{\Delta^2} + \frac{
\left[ P^2 -K^2+\Delta^2\right]\left[P^2-K^2-\Delta^2\right]}{\Delta^2
\surd \left[ (P-K)^2 +\Delta^2 \right] \surd \left[ (P+K)^2 +\Delta^2
\right]} \right\} .
\end{equation}
In fact this approximation is also good for the normal phase $M=0$.
In the $P\longrightarrow 0$ limit,
\begin{equation}
I(P=0,K) = - \frac{4\pi (\Delta^2 -K^2)}{(\Delta^2+K^2)^2} .
\end{equation}
The effective ultra violet cut--off means that the momentum $K$ must
satisfy
\[
K \in \left[ \surd(\alpha/\beta) ,\, \alpha \right].
\]
Then in an intermediate momentum range
\[
\surd(\beta\alpha) \sim \alpha/\varepsilon \gtrsim 1,
\]
and \( K\in \left[ \surd (\alpha/\beta) , \alpha\right]\) to within
\({\mathcal O}(\alpha)\).

Hence the wavefunction renormalization is given by
\begin{eqnarray}
A(P=0, \beta) &\simeq& 1 - \frac{1}{4\pi N_{\rm f}}
\int_{\surd(\alpha/\beta)}^\alpha dK \; \frac{1}{K} \tanh\left(
\frac{\beta K}{2} \right) \simeq \nonumber\\
&\simeq& 1- \frac{1}{4\pi N_{\rm f}}
\int_{\surd(\alpha\beta)/2}^{\alpha\beta/2} dx\; \frac{\tanh x}{x} .
\end{eqnarray}
If the temperature is very low, $\alpha\beta \gg 1$:
\begin{equation} \label{approxA}
A(P=0, \beta) \simeq 1 - \frac{1}{8\pi N_{\rm f}} \ln (\alpha\beta) .
\end{equation}
For the region $\alpha\beta \gtrsim 1$ the integrals can be evaluated
numerically and it is found that equation (\ref{approxA}) is
reasonably accurate:
\begin{eqnarray} \label{stability}
\alpha\beta \gtrsim 5 &\longrightarrow& {\rm Result\ is\ within\ 10\%\
of\ (\ref{approxA})};\nonumber\\
\alpha\beta \gtrsim 10 &\longrightarrow& {\rm Result\ is\ essentially\
exact.}
\end{eqnarray}

The resistivity can be computed from equation (\ref{resisteq}) using
(\ref{approxA}) for the finite temperature wavefunction
renormalization, with the result
\begin{equation} \label{scalingexpt}
\rho \propto T^{1- 1/4\pi N_{\rm f}}  \qquad (n=1) .
\end{equation}
Note that the critical exponent is only approximate.  Above it has
been calculated by working in a specific gauge (the Landau gauge:
$\xi=0$) and by adopting a specific {\sl ansatz\/} for the vertex,
equation (\ref{webb}).  However, since the resistivity is a physical
gauge invariant quantity, the scaling (\ref{scalingexpt}) should be
gauge--independent.  This gauge--independence should be checked
explicitly by either performing the computations in many different
gauges upon selecting appropriate vertex {\sl ans\"{a}tze}, or
following the alternative approach to the running in theory space
based on a Dyson--Schwinger treatment for the amputated vertex
described in section \ref{papsec}.  This method has the advantage that
it avoids the problem of selecting a vertex {\sl ansatze\/} satisfying
the Ward--Takahashi identity, for it determines the vertex directly.

From the relation (\ref{stability}) it is apparent that this scaling
is stable for a large temperature range, in agreement with the
experimental situation \cite{varma89,malinowski97}.  Note that the
scaling exponent in relation (\ref{scalingexpt}) is {\em less} than
one (by a small amount, at least within the $1/N_{\rm f}$ treatment)
in the gauge field theory model.  Such a scaling seems to characterize
the normal phase of the high temperature cuprates \cite{malinowski97}.
In the actual experimental observations the scaling depends on the
doping concentration, $\delta$, in such a way that the exponent
diminishes linearly with $\delta$.  In the simplified field--theoretic
analysis above a precise connexion between this exponent and the
doping concentration cannot be made, and therefore this dependence
cannot be verified directly.  However, from the connexion between
QED${}_3$ and underlying statistical models \cite{doreycombo,dorey91},
a dependence on $\delta$ in the exponent in equation
(\ref{scalingexpt}) is to be expected from the following general
argument: in QED${}_3$, $1/N_{\rm f} \sim e^2/\alpha$, and from the
analysis in reference \cite{dorey91} the coupling $e^2$ depends on the
doping concentration.  Since a detailed microscopic model is lacking
at present, a phenomenological approach may be adopted to try to infer
the doping dependence of the gauge coupling by comparing the
resistivity curves fitted from equation (\ref{scalingexpt}) with the
experimental ones \cite{malinowski97}.

\subsection{Comparison with other approaches.}

The scaling law (\ref{scalingexpt}) is an interesting prediction of
the gauge spin--charge separation approach to high temperature
superconductivity, however it may not be unique to this model.
Indeed, such scaling can also be obtained in different approaches
\cite{nayak94,byczuk}.  In some of them \cite{nayak94} the
non--trivial infra red fixed point structure plays a r\^{o}le in
inducing the deviations from the linear resistivity in analogy with
equation (\ref{scalingexpt}).  In other treatments \cite{byczuk}, a
phenomenological approach to Luttinger and Fermi liquids has been
adopted by combining spin--charge separation with the anomalous
scaling of a Luttinger liquid in postulating the following
unconventional form for the one electron retarded Green's function
\cite{byczuk}:
\begin{equation} \label{branchcuts}
G_{\rm R} (\vec{k},\omega) \propto \frac{1}{\left(\omega -
\epsilon^c_{\vec{k}} + i 0^+ \right)^\beta \left(\omega -
\epsilon^s_{\vec{k}} + i0^+ \right)^\gamma },
\end{equation}
where appropriate phase factors have been left out for simplicity.
For low energy excitations near the Fermi surface
\[
\epsilon^{c,s}_{\vec{k}}= v^{c,s}_{\rm F} |\vec{k} -
\vec{P}_{\rm F}| ,
\]
the indices $c,s$ respectively denoting charge and spin degrees of
freedom.  In this approach spin--charge separation arises because of
the different propagation velocities $v^{c,s}_{\rm F}$ in a direct
generalization of the one dimensional case.  What the authors of
\cite{byczuk} argue is that upon requiring time--reversal invariance
(in the absence of external fields) the exponents in the Green's
function (\ref{branchcuts}) are given by
\[
\beta=\gamma=\frac{1}{2} - \chi.
\]
The Fermi liquid theory corresponds to $\chi=0$ and $v^c_{\rm F} =
v^s_{\rm F}$, in which case the Green's functions (\ref{branchcuts})
have only single pole excitations corresponding to the electronic
degrees of freedom.  In the $\chi\neq0$ and $v^c_{\rm F} \neq v^s_{\rm
F}$ case there are branch cuts corresponding to spinon and holon
quasiparticle excitations.  The presence of the $\chi\neq0$ exponent
leads to a scaling of the direct current resistivity $\rho$ in the
normal phase of the high--T${}_{\rm c}$ systems of the form
\begin{equation} \label{by}
\rho \sim T^{1 - 4\chi} .
\end{equation}
The authors of \cite{byczuk} fit this scaling with the experimental
data \cite{malinowski97} by postulating $\chi>0$ and assuming a doping
dependence.  Although the method above made use of a different
microscopic treatment of spin--charge separation (equation
(\ref{spinchargeeq})) the similarity of the scaling laws
(\ref{scalingexpt}) and (\ref{by}) encourages further studies in order
to try to establish a connexion between the two approaches.

\section{Conclusions.} \label{conc}
\setcounter{equation}{0}

In these lectures a renormalization group approach to the effective
field theory of condensed matter systems has been discussed.  The
approach proves to be extremely powerful insofar as it yields useful
information on the universal behaviour of apparently different
physical systems.  The basic properties of the renormalization group
have been described briefly, and the various effective field theory
interactions have been classified according to their renormalization
group scaling properties.  BCS superconducting instabilities were
studied by means of an exercise for the reader: they were described in
terms of relevant renormalization group operators driving the theory
away from the trivial infra red fixed point (Landau's Fermi liquid).
Particular attention was paid to discussing quasiparticles, which are
linearized excitations about the Fermi surface which exhibit
appropriate scaling under the renormalization group.  Such
linearizations led naturally to the concept of an effective fermion
``flavour number,'' which is the (large) number of excitations near
the Landau fixed point in condensed matter systems with large Fermi
surfaces.  In such cases, the area of the Fermi surface may be
identified (roughly) with the flavour number.  Such
large--$N_{\rm f}$ effective theories have a running flavour number
due to the dependence in $N_{\rm f}$ on the cut--off scale in the
theory.

In the second part of the lectures an interesting application of the
above ideas, in particular the running flavour number, was discussed
via the study of infra red structure in strongly coupled U(1) gauge
theory with relativistic massless fermions.  Physical arguments were
given as to why such models may be relevant for the physics of high
temperature superconducting cuprates.

The (non--perturbative) infra red structure of QED${}_3$ has been
studied extensively using a renormalization group improved
Dyson--Schwinger approach, resummed to leading order in $1/N_{\rm f}$.
Self consistent solutions have been found in Landau and non--local
gauges, and the importance of the infra red cut--off has been
discussed.  The non--trivial infra red fixed point leads to very slow
running of the coupling, which may have important consequences for the
physics of systems described by these models ({e.g.}\ the size of the
gap and the associated coherence length in superconductors).  The
computation of some critical exponents using the non--local gauge and
Landau--Khalatnikov transformation was also discussed.

Discontinuities arising from the removal of covariant infra red
cut--offs and related Landau damping type effects in finite
temperature theories have also been discussed.  The connexion leads to
natural predictions for the resistivity in gauge field mediated
models.  Gauge interactions lead to a resistivity linear in the
temperature, and the slow running coupling (affecting the system
through the wavefunction renormalization) alters this behaviour with
${\mathcal O}(1/N_{\rm f})$ corrections in the exponent.  It is
possible that these corrections are experimentally testable.  In
fact, recent experimental data seem consistent with slight deviations
from linear $T$ behaviour in the direct current resistivity
\cite{malinowski97}.

The fundamental ideas which have been covered are as follows:

\begin{itemize}
\item Renormalization group analysis for U(1) gauge theories in
a large--$N_{\rm f}$ framework $\Longrightarrow$ Dyson--Schwinger
analysis improved by the renormalization group.
\item Connexion with Landau--Fermi theory:  an increase in the area of
the Fermi surface corresponds to an increase in the number of
flavours, and hence eventually the large--$N_{\rm f}$ limit.
\item Approach the superconducting (chiral symmetry broken) phase from
the normal phase.
\item Use the renormalization group to investigate the evolution of
Fermi surfaces in planar systems: reduction to a small momentum range
in the deep infra red (in an appropriate doping regime), which admits
a {\em relativistic} effective field theory description.
\item It is found that the effective $N_{\rm f}$ (in the large--$N_{\rm
f}$ regime) is renormalized, and runs with the scale:  this generates a
running in theory space.
\item This slow running of $N_{\rm f}$ is an exclusive feature of U(1)
gauge theories. The gauge interactions have a tendency to reduce the
size of the Fermi surface, and hence lead to non--Fermi liquid
behaviour.
\end{itemize}

\section*{Acknowledgments}

N.E.M. would like to thank J. Spa{\l}ek and the other organizers of
the XXXVIII Cracow School of Theoretical Physics for inviting him to
lecture at the School, and for providing a stimulating atmosphere
during the meeting.  We would also like to acknowledge helpful
discussions with I.J.R. Aitchison, N. Andrei, C. Hooley, M. Lavagna,
J. Papavassiliou, S. Sarkar 
and J. Spa{\l}ek.  The work of N.E.M. is supported by a
P.P.A.R.C. (U.K.) Advanced Research Fellowship.  A.C.--S. would like
to thank P.P.A.R.C. (U.K.)  for a research studentship (number
96314661).

%\bibliography{/home/wytham/danton/bibtex/references}

\end{document}